\documentclass[a4paper]{article}

\usepackage[english]{babel}
\usepackage[utf8x]{inputenc}
\usepackage[T1]{fontenc}

\usepackage[a4paper,top=3cm,bottom=2cm,left=3cm,right=3cm,marginparwidth=1.75cm]{geometry}

\usepackage{amsmath}
\usepackage{rotating, graphicx}
\usepackage[colorinlistoftodos]{todonotes}
\usepackage[colorlinks=true, allcolors=blue]{hyperref}
 \def\nnu{{{\mathrm N}_{\nu}}}
   \def\epem{\rm{e^+e^-}}
 
  \def\mz{{{{\mathrm M}}}_{{{\mathrm Z}}}}
  \def\MMZ{{{{\mathrm M}}}_{{{\mathrm Z}}}^2}
  
     \def\ra{\rightarrow}

  \def\eqref#1{(\ref{#1})}

   \def\su2{{{\mathrm{ SU(2)}}}_{{\mathrm L}} \times{{\mathrm U(1)}}}

  \def\sgaf{g_{Af}}
  \def\sgvf{g_{Vf}}

  \def\as{\alpha_{s}}

  \def\mf2{m_f^2}
  
   \def\mz{{{{\mathrm M}}}_{{{\mathrm Z}}}}

  \def\MMZ{{{{\mathrm M}}}_{{{\mathrm Z}}}^2}

  \def\nnu{{{\mathrm N}_{\nu}}}

  \def\gamnu{{\Gamma}_{\nu}}

  \def\gamlep{{\Gamma}_{\ell}}
  
\def\GH{{\Gamma}_{\mathrm{ had}}}

  \def\GL{{\Gamma}_{\ell}}

  \def\Rl{{{\mathrm R}_{\ell}}}

  \def\s0h{{\sigma}^{{\mathrm{ peak,0}}}_{{\mathrm{ had}} }}

  \def\gamz{{\Gamma}_{{{\mathrm Z }}}}
  \def\Gz{{\Gamma}_{{{\mathrm Z}}}}
  \def\GZ{{\Gamma}_{{{\mathrm Z}}}}
  
  \def\sef{{ \sin^{2} {\theta }_{\mathrm w}^{{{\mathrm {eff}}}}}}

\newcommand{\MZ}      {m_{\mathrm{Z}}}

\title{The third family of neutrinos}
\author{Alain Blondel}

\begin{document}
\maketitle

\begin{abstract}
This paper retraces the 24 years starting with the appearance of the symbol "$\nu_{\tau}$" in 1977, until the observation of tau neutrino interactions with matter in 2000. The fact that the neutral particle present in tau decays was a neutrino was demonstrated by 1979; its existence as the third neutrino $\nu_{\tau}$, iso-spin partner of the tau lepton, was definitely established in 1981-1986; it was demonstrated that the number of light active neutrinos is closed with the known ones ($\nu_e, \nu_{\mu},\nu_{\tau}$) in 1989; the $\nu_{\tau}$ properties had been precisely determined in $e^+e^-$ and $p\bar{p}$ collider experiments. \\
\\
{\sl Contribution to the Conference on the History of the Neutrino, Paris, 5-7 September 2018} 
\end{abstract}

\section{Summary}

The initial tau lepton observation, in 1975 at SPEAR, was based on its decay into neutrinos together with electrons or muons~\cite{Perl:1975bf}~\cite{Feldman:1976fm}~\cite{Perl:1977se}. It took a few more years to identify "$\nu_{\tau}$", the neutral, weakly (V-A)-interacting spin $1/2$ particle~\cite{Bacino:1979fz}~\cite{Kirkby:1979pv}, 
observed also in two-body decays 
such as $\tau \rightarrow \pi {\nu}$~\cite{BarbaroGaltieri:1977ti}\cite{Brandelik:1977xz}\cite{Blocker:1981mc}, to be the isospin partner of the tau, $\nu_\tau$, distinct from $\nu_e$ and $\nu_{\mu}$. In 1981~\cite{Feldman:1981nv} the distinct existence of the tau neutrino could be established with moderate statistical significance, from the combination of non-observation of tau leptons in high energy neutrino beams~\cite{Cnops:1977zc}~\cite{Fritze:1980un} with the first measurements of the tau lifetime~\cite{Feldman:1981md}; PDG82~\cite{Roos:1982sd,PDGarchives} thus listed $\nu_{\tau}$ as established particle. High significance was gained from i) more precise measurement of the tau lifetime by several experiments; ii) the observation of $ W \rightarrow \tau  {\nu}_{\tau}$ ~\cite{Vuillemin:1985tc}\cite{Albajar:1986fn} with the same rate as $ W \rightarrow e   {\nu}_{e}$  and $ W \rightarrow \mu  {\nu}_{\mu}$~\cite{Albajar:1988ka}; and iii) the E531 emulsion neutrino experiment at Fermilab~\cite{Ushida:1986zn}. By the start of LEP in 1989, the tau-neutrino was a well-established particle.\\

Meanwhile, the impact of the number of light, active neutrinos on cosmology and astrophysics was appreciated, and the question of further families of neutrinos had become a critical input. A number of measurements, from astrophysics to collider physics, pointed to a number of light active neutrinos, $N_\nu$, situated between 2 and 4~\cite{Denegri:1989if}. The number of light, active neutrinos was presented by the MarkII experiment at the SLC~\cite{Abrams:1989yk} and from the first three weeks of data taking by the LEP collaborations L3~\cite{Adeva:1989mn}, ALEPH~\cite{Decamp:1989tu}, OPAL~\cite{Akrawy:1989pi}and DELPHI~\cite{Aarnio:1989tv}, in a grand seminar at CERN on 13 October 1989. 
The combined number $N_\nu = 3.12 \pm 0.19$~\cite{blondel1991} demonstrated that there are no further types of light active neutrinos than the three ($\nu_e,\nu_{\mu}, \nu_{\tau} $) that were already known; the latest LEP value is $N_\nu = 2.984 \pm 0.008$ ~\cite{ALEPH:2005ab}. \\

A million of tau pairs were produced in Z decays by the LEP collaborations. The tau lifetime and branching ratios were determined with much increased precision, establishing the equality of the $\tau -- {\nu}_{\tau}$ coupling with those of the other lepton doublets~\cite{Charlton:1993as}, eventually reaching a precision of a few permil~\cite{lepton-universality-HFLAV-2017}. At LEP2, tens of thousands of W pair events were detected with a clear event-by-event kinematic reconstruction allowing determination of all three  W lepton-neutrino couplings at the few percent level\cite{Schael:2013ita}. In 2000 the DONUT collaboration, in a dedicated beam containing ~5\% of the, by then well known, tau neutrino, made the first observation of its charged current interactions with matter~\cite{Nakamura:1999dp,Kodama:2000mp,Kodama:2007aa}.  
\\

This history leads to a reflexion on the scientific merit of the words {\em direct}, {\em indirect} and {\em discovery}
that have sprinkled the history of the third and last light active neutrino. The words {\em direct} and {\em indirect} are quite subjective, and  used in different ways for different problems; more quantitative gauges, i) the statistical (and systematic) significance, and ii) the possible model-dependence, should be preferred. As for "discovery", its definition is "the act of finding something that was not known before". In this definition the moment where the tau-neutrino was discovered (found to exist as a new particle) can be situated between 1981 at the earliest, with the first tau lifetime estimate; and at the latest and definitely in 1986, after it was shown  that, while the tau lepton couples with full weak interaction strength to its neutrino, electron and muon neutrinos do not couple to it significantly.


\section{The first appearance of the symbol "$\nu_\tau$"}

The history of the third family neutrino begins with the discovery of the third family lepton in the years 1975 to 1977 by M. Perl and the SLAC-LBL magnetic spectrometer collaboration, now known as MARK-I. A personal recollection of this time was given by G. Feldman in "the tale of three papers"~\cite {Feldman:1992vk} which contains many references of the time, and a step by step review of the physics arguments. 

\begin{figure}[h]
\centering
\includegraphics[width=0.40\textwidth]{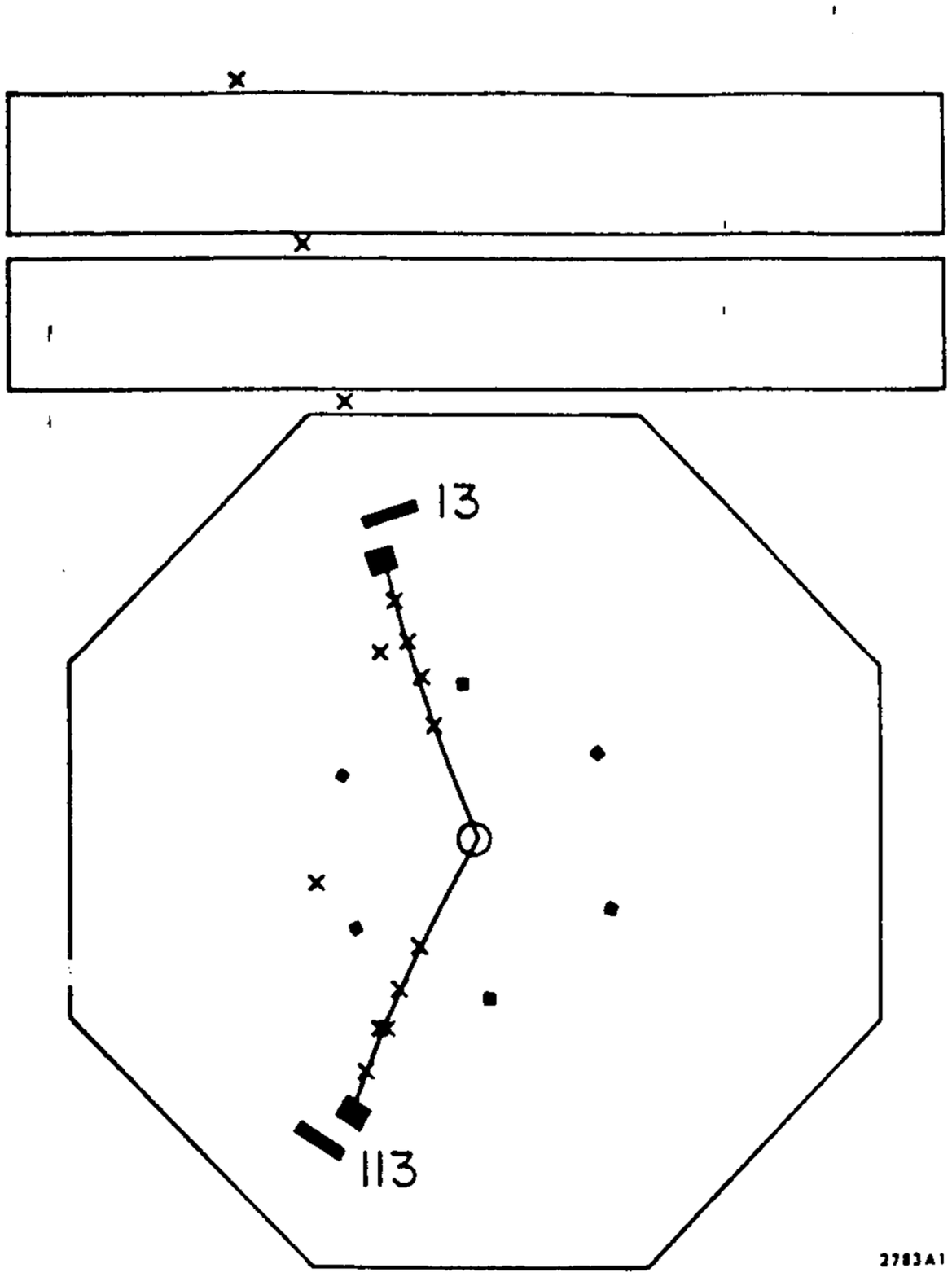}
\includegraphics[width=0.40\textwidth]{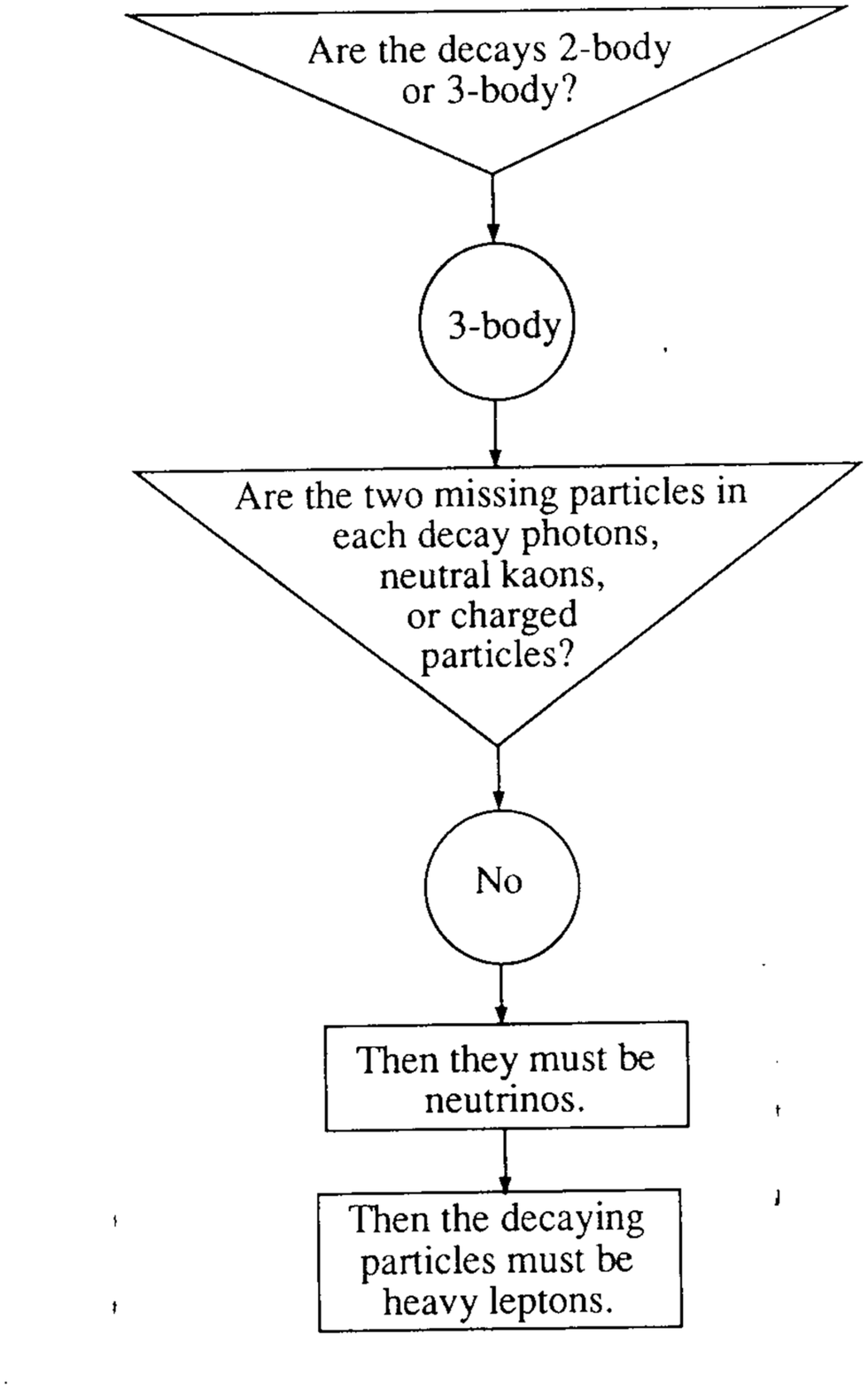}
\caption{\label{fig:tau-e-mu} \small  Left: One of the first e-$\mu$ events with particle identification in the muon 'tower' and in the shower counters, as shown in the Standford 1975 Lepton Photon Symposium. Right: the demonstration that the new particle is a lepton is based on the presence of neutrinos in the final state.}
\end{figure}

The heavy lepton search was based on the sequential lepton scheme, 
\begin{equation}~
\label{eq:heavy-lepton}
e^+ e^- \rightarrow L^+ L^- ; L^+ \rightarrow \mu^+ \nu \nu {\rm ~and~} L^- \rightarrow e^+ \nu \nu
\end{equation}
and combinations of charge and lepton flavour. The production and decay properties of such particles had been described by P.Tsai~\cite{Tsai:1971vv}. The characteristic signatures are two leptons of different flavours and, importantly, large amounts of missing energy and momentum going into neutrinos, as is clearly visible already on one of the first event displays, Fig.~\ref{fig:tau-e-mu}, left. We will here concentrate on the neutrinos. While the first MARKI paper~\cite{Perl:1975bf} focuses on the appearance of anomalous e-$\mu$ events, the second one~\cite{Perl:1976rz} is mostly concerned with the demonstration of the fact that the new particle is a lepton. Here the presence of neutrinos in the final state is essential; they were identified by the absence in the anomalous e-$\mu$ events of photons or other electromagnetically decaying particles ($\pi^0, \eta$ etc.), and of neutral hadrons, neutrons or ${\rm K_0, \Lambda}$.

\begin{figure}[ht!]

\centering
\vspace{-8cm}
\includegraphics[width=0.95\textwidth]{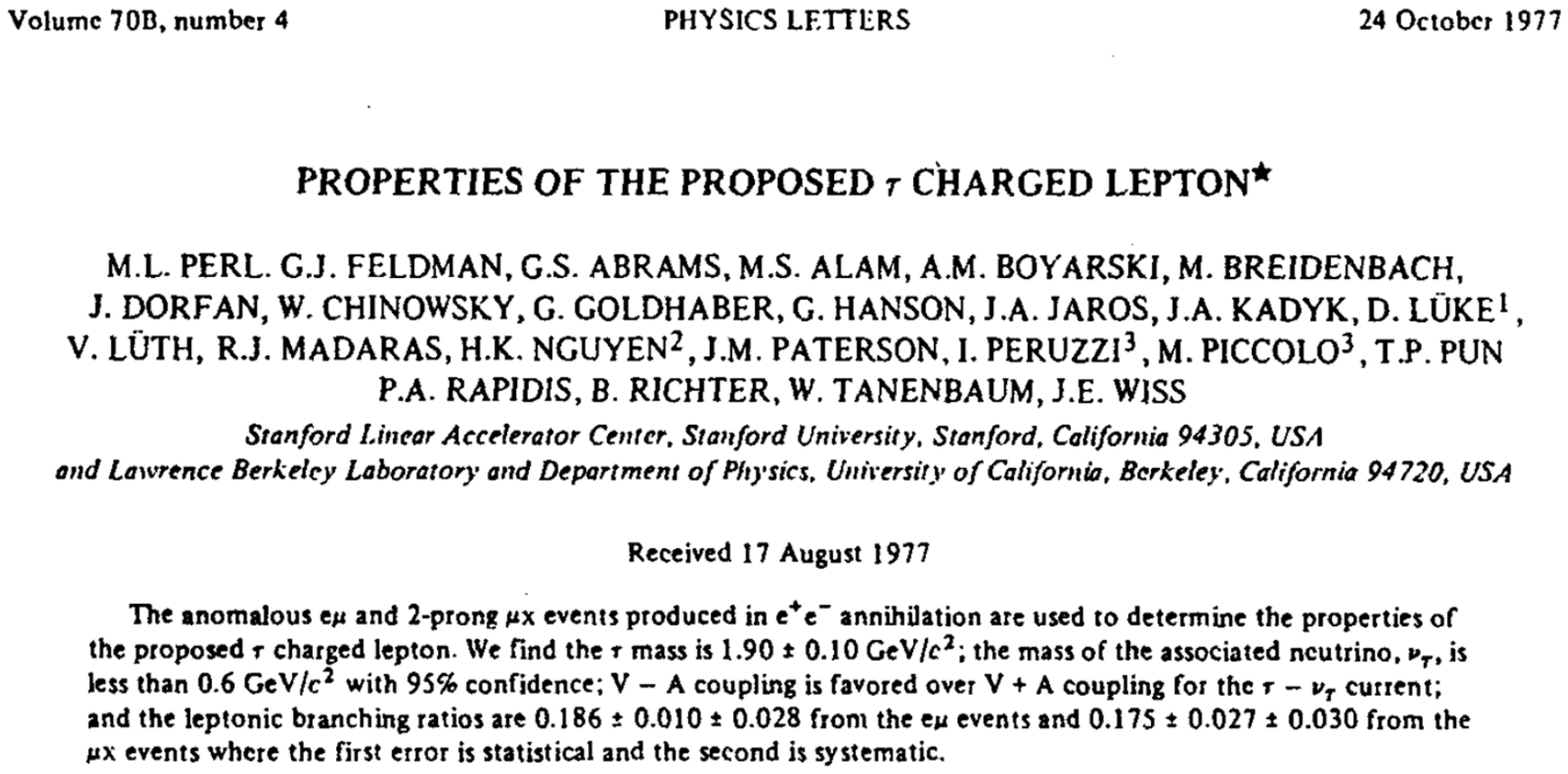}
\vspace{-7cm}
\caption{\label{fig:nu-tau-appears} \small  The first appearance of the symbol $\nu_\tau$ ~\cite{Perl:1977se} }
\end{figure}

The new particle being established, it was given the name $\tau$, as explained in~\cite {Feldman:1992vk}: {\sl it had to be Greek like the muon, so $\tau$, as in $\tau \rho \iota \tau o \nu$ (third), was chosen} -- among other possibilities. This name appears in the third paper~\cite{Perl:1977se}, Fig~\ref{fig:nu-tau-appears} 'Properties of the proposed $\tau$ charged lepton'. Together with it, without much further ado, appears the symbol "$\nu_\tau$" as in "{\sl We find the $\tau$ mass is $1.90 \pm 0.10$ GeV/$c^2$ ; the  mass of the associated neutrino $\nu_\tau$ is less than 0.6 GeV/$c^2$ with 95\% confidence; V-A coupling is favored over V+A coupling for the $\tau - \nu_\tau$ current;}" etc.  

The fact that the $\tau$ lepton behaved as expected from a sequential heavy lepton could certainly justify the presumption that the additional neutral particle, that had to be there in the three body leptonic decays $\tau \rightarrow e/\mu \bar{\nu_{e/\mu}} \nu $  in addition to the anti-neutrino of the same flavour as the tagging lepton ($e/\mu$) would be the $\tau$'s weak isospin partner, $\nu_\tau$. However this was not a fool-proof demonstration, and there could be other hypotheses. This caused a general and long lasting impression that the existence of the tau neutrino had been  'inferred' rather than properly demonstrated.  
We will see that a solid demonstration would take place from then to 1986.   

\begin{figure}[htb!]
\centering
\includegraphics[width=0.48\textwidth]{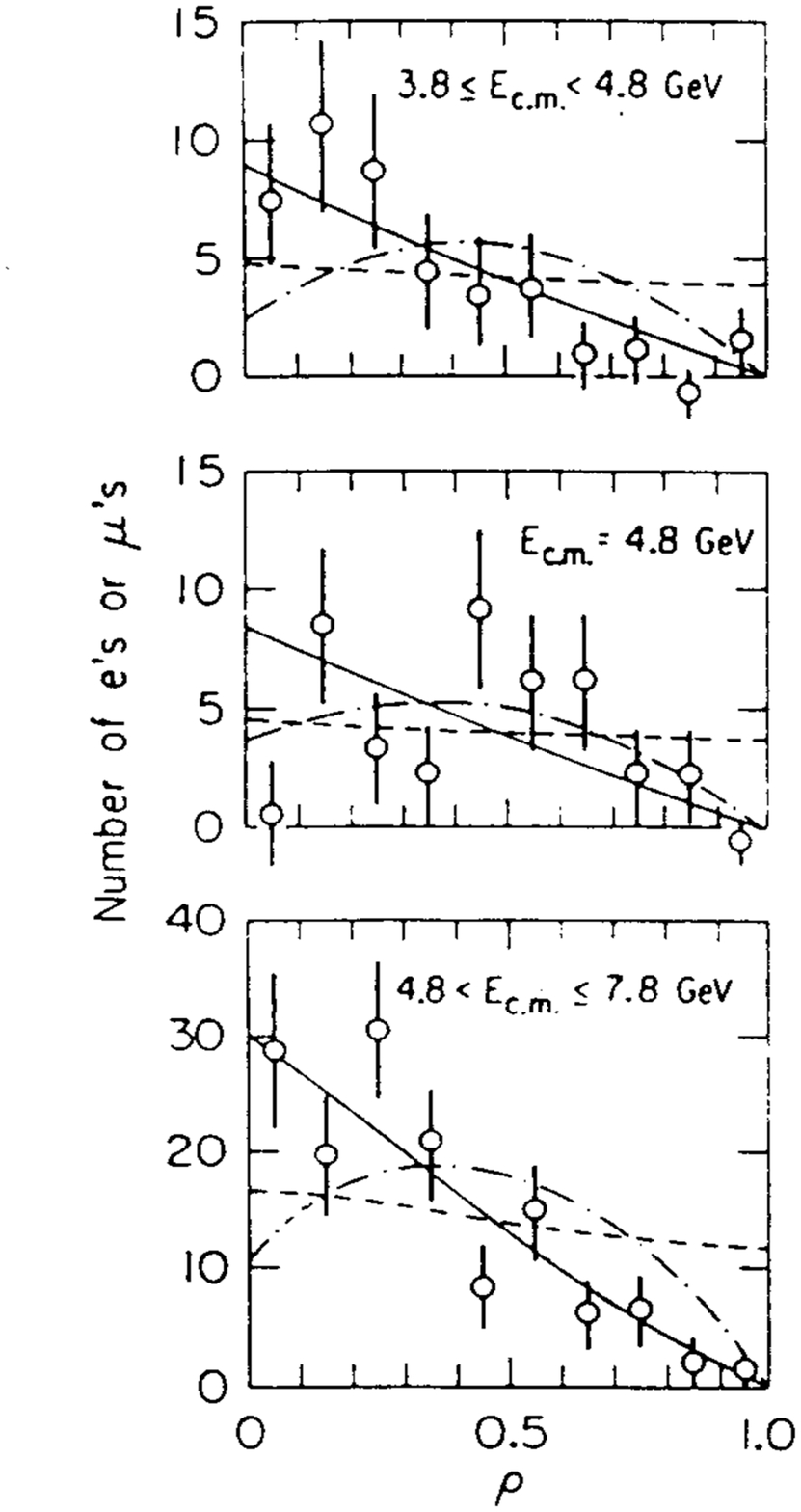}
\includegraphics[width=0.48\textwidth]{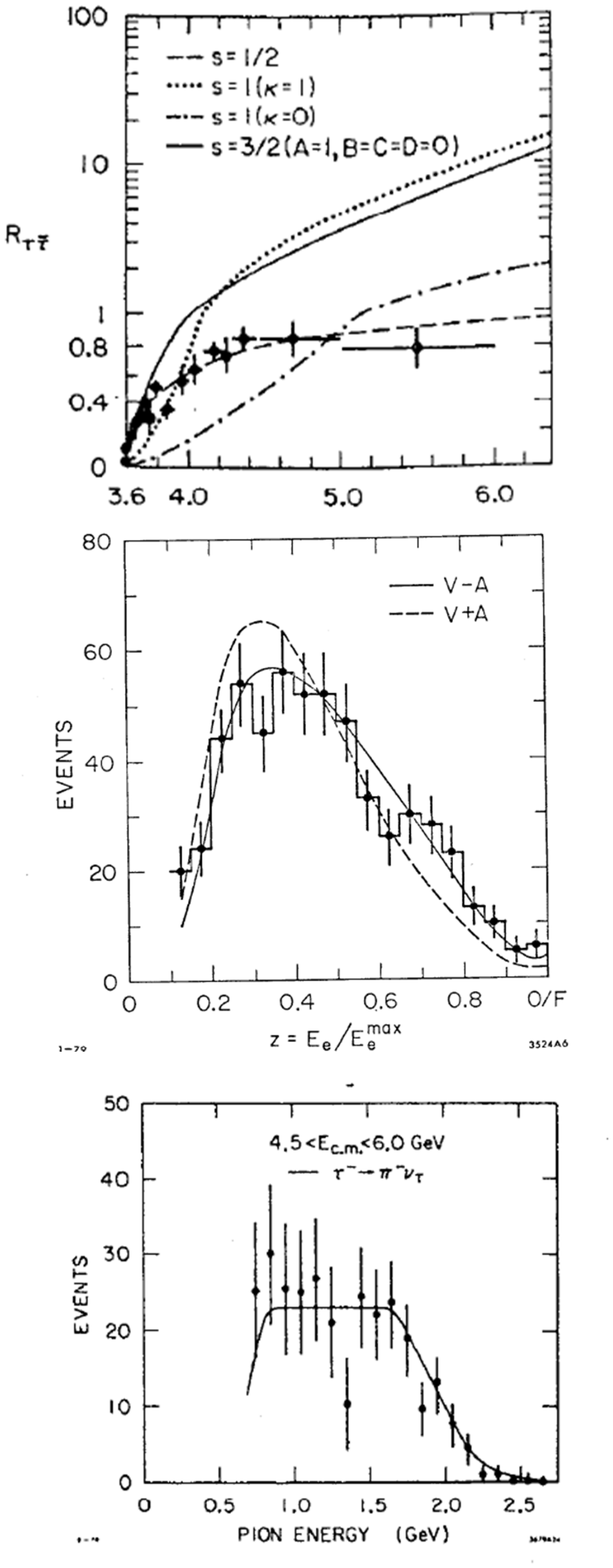}
\caption{\label{fig:taudecay-properties} \small  Left: normalized spectra (${\rm \rho=p/p_{beam}}$) of the leptons in the anomalous e-$\mu$ events, showing that the  V-A coupling is favored over V+A coupling for the $\tau - \nu_\tau$ current. Right, top: the tau cross-section from 3.5  to 6.5 GeV ${\rm e^+e^-}$ centre-of-mass energy, demonstrating that the new particle is a spin 1/2 lepton~\cite{Kirkby:1979pv}; middle: The full lepton spectrum consistent with V-A in DELCO~\cite{Kirkby:1979pv}; bottom: the momentum spectrum of pions in the decay $\tau \rightarrow \pi \nu$, showing consistency with the neutrino has a mass < 250 MeV/$c^2$.   }
\end{figure}

\section{"$\nu_\tau$" is a neutrino}
During the following four years (1977-1981), measurements of  cross-section and decays by MARKI, MARKII, DELCO, at SPEAR, PLUTO and DASP at DORIS, (individual references can be found in the PDG, 1982 edition~\cite{Roos:1982sd} ) were performed. They readily showed the following, also summarized in Fig.~\ref{fig:taudecay-properties}.    
\begin{itemize}
\item The tau decays into leptons and two neutrinos and the decay is V-A
\cite{Perl:1977se,Kirkby:1979pv} 
\item The tau pair cross section raises sharply from the threshold around 3.5 GeV and reaches quickly the same level as that of muon pairs, 
demonstrating that the tau lepton itself is a spin ½ particle~\cite{Kirkby:1979pv}.     
\item It was reported already in 1977~\cite{BarbaroGaltieri:1977ti} \cite{Brandelik:1977xz}\cite{Blocker:1981mc}, that the tau decays into hadron and one neutrino, in particular $\tau \rightarrow \pi \nu$, with a two body decay -- this is a very direct proof of the existence of a single  spin ½ neutral particle, the spin 1/2 tau cannot decay into a spin 0 pion and a spin 3/2 (or more) particle~\cite{Kirkby:1979pv}.  
\item The tau was also observed to decay in ${\mathrm \rho \nu, K*\nu, A_1\nu}$ etc… in a way that is consistent with the weak hadronic current. In particular the K*/$\rho$ ratio is consistent with the Cabbibo angle, a trademark of the weak hadronic current. 
\item Further studies of the $\tau \rightarrow \pi \nu$ decay~\cite{Blocker:1981mc} showed excellent agreement with the two body hypothesis with an upper limit on the mass of the "$\nu_\tau$" of 250 MeV/$c^2$.   

\end{itemize}

All this established the "$\nu_\tau$" particle as a spin 1/2 neutral lepton interacting by V-A interactions, and with a mass consistent with zero within the experimental resolution. This is what we call a neutrino. 

\begin{figure}[htb!]
\centering
\vspace{-7cm}
\includegraphics[width=0.95\textwidth]{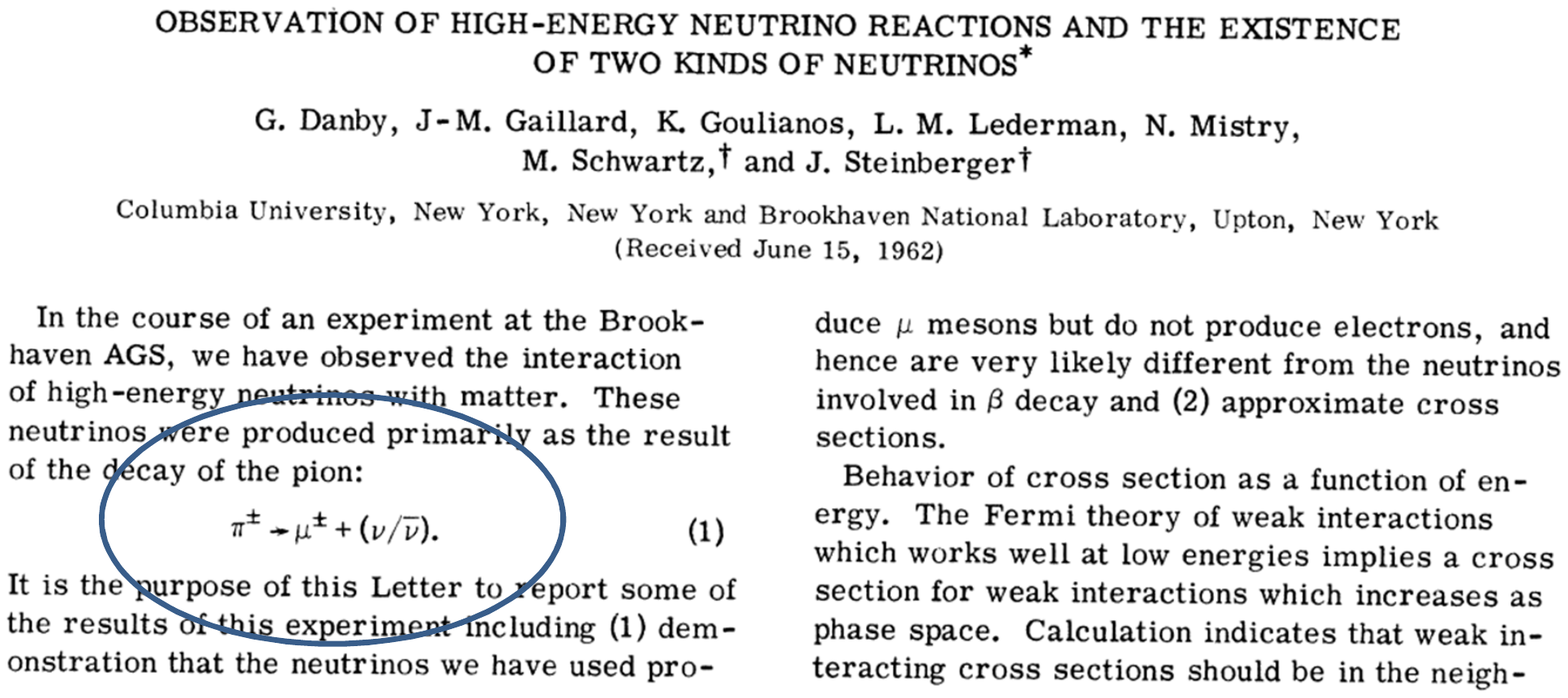}
\vspace{-6cm}
\caption{\label{fig:Brookhaven-numu} \small  The first page of the Brookhaven neutrino experiment of 1962~\cite{Danby:1962nd}. }
\end{figure}

The situation at that point was comparable to that preceding the 1962 Brookhaven neutrino experiment ~\cite{Danby:1962nd}, for which Lederman, Schwartz and Steinberger were awarded the Nobel Prize in 1988. There was no question in the late fifties that there was a neutrino produced in the two body decay $\pi^\pm\rightarrow \mu^\pm (\nu/\bar{\nu})$, see Fig.\ref{fig:Brookhaven-numu}. The question was whether this neutrino was the same as in beta-decay ($\nu_e$) or whether it was a neutrino of a new kind. Similarly for the "$\nu_\tau$" in the late seventies, the question was not whether there was a neutrino in e.g. $\tau \rightarrow \pi \nu$  decays, but whether it was one of a new kind. 

There was however a very significant difference: producing a 99\% pure beam of "$\nu_\tau$" would be extremely difficult and it was out of reach at the time -- and still is difficult. So that another method had to be found, to demonstrate that the "$\nu_\tau$" is a new particle and not $\nu_e,\nu_\mu $ or some mix or superposition thereof.

\section{A new neutrino? }
Is it possible that the  "$\nu_\tau$" would not be  the weak isospin partner of the tau? Yes.  

At the same epoch, (1977-1981) the b-quark had been discovered, decaying into charm and not into a new third generation quark, because the top quark is heavier than the b quark. As a consequence the b decay is suppressed by the CKM element («mixing angle»)   ${\rm V_{cb}}$, and  the b-meson lifetime was indeed found to be much longer than would be expected given its mass\cite{Fernandez:1983az, Lockyer:1983ev} . 

The same thing could happen with the tau lepton, for instance if $\nu_\tau$ were heavier than the tau lepton. Then what had been initially called "$\nu_\tau$" would be indeed a $\nu_e$  or $\nu_\mu $ or some mix or superposition thereof. Parethetically, it had been also proposed that the "$\nu_\tau$" would be either ${\bar \nu}_e$ or ${\bar \nu}_\mu$; however this would have affected either the corresponding leptonic partial widths, challenging their already observed equality and/or their ralation to the $\pi \nu$ decay mode~\cite{petcov2018}; this would have eventually been also ruled out later, once the tau neutrino helicity had been measured. In the following we concentrate on the hypothesis that the "$\nu_\tau$" would be indeed a $\nu_e$  or $\nu_\mu $ or some mix or superposition thereof. 

This hypothesis would imply that 
i) the tau lifetime would be longer than predicted if a 100\% 
$\tau - \nu_\tau$ is assumed.
ii) because the tau would couple to  $\nu_e$  and/or $\nu_\mu $, 
taus could be produced in neutrino beams of sufficient energy; the kinematic threshold being 3.5 GeV neutrino energy, beams at the CERN SPS or Fermilab Tevatron, with typical energies of 20 GeV or more, would do it.

To demonstrate that the tau neutrino was a the weak isospin partner of the tau one should therefore demonstrate both of the following.
\begin{enumerate}
\item Show that the coupling of the tau to its neutrino has the full weak interaction strength -- this could be done by measuring the tau lifetime, and from this and from the branching ratio into $e \nu\nu$ extract the  $\tau \rightarrow e \nu\nu$ partial width and, {\sl mutandis mutandi}, show that the coupling strength is the same as that of the muon; alternatively show that the  W boson decays into $W\rightarrow \tau \nu$ with the same rate as $W\rightarrow e \nu$  and $W\rightarrow \mu \nu$.   

\item Show  that  neither $\nu_e$  nor $\nu_\mu $  couple to the tau.
\end{enumerate}

The tau lifetime measurements can be made in $e^+e^-$ experiments with sufficient beam energy and if the tracking device can measaure a significantly positive impact parameter for the tau lepton decay products.  Experiments at the time did not have the precise silicon vertex detectors that are now customary in collider experiments. In addition, in the energy range of SPEAR and DORIS (from threshold to 7 GeV) the tau lepton velocity was too small to observe a significant life-time. The first measurements were made at the higher energy $e^+e^-$ colliders PETRA at DESY and PEP at SLAC, with TASSO~\cite{Brandelik:1980ga} from 12 to 31.6 GeV ECM, and  MARKII~\cite{Feldman:1981md} at 29 GeV ECM respectively. The detectors were  drift chambers of 200-300 micron r-$\phi$ point resolution situated 10-15 cm from the  beam axis. In 1981 the existing measurements averaged to $(4.6\pm1.9)\times 10^{-13} s$, for the lifetime and $16.2 \pm 1.0$\% for the $e \nu\nu$ branching ratio~\cite{Roos:1982sd}.    

The search for production of tau leptons (and more generally heavy leptons) in neutrino interactions was already started in 1977, in the 15-foot cryogenic bubble chamber at Fermilab, filled with liguid neon~\cite{Cnops:1977zc}. The liquid neon having a radiation length of 40cm and interaction length of 1m, all in a magnetic field of 3T, it was readily possible to identify prompt electrons and prompt muons from hadrons. The tau production signal would be production of an excess of prompt electrons at high energies (over the intrinsic contamination from the beam).  In a sample of 27600 charged current $\nu_\mu $ interactions, 187 $e^-$ and 28 $e^+$ were observed against an expectation of $215 \pm 60 \nu_e $  and $23 \pm 8 \bar {\nu_e} $ interactions from the beam calculations. Hence an upper limit on coupling strength of  $\nu_\mu $  to the tau must be less than
0.025 of the coupling strength of $\nu_\mu $ to the muon.

\begin{figure}[htb!]
\centering
\vspace{-7cm}
\includegraphics[width=0.95\textwidth]{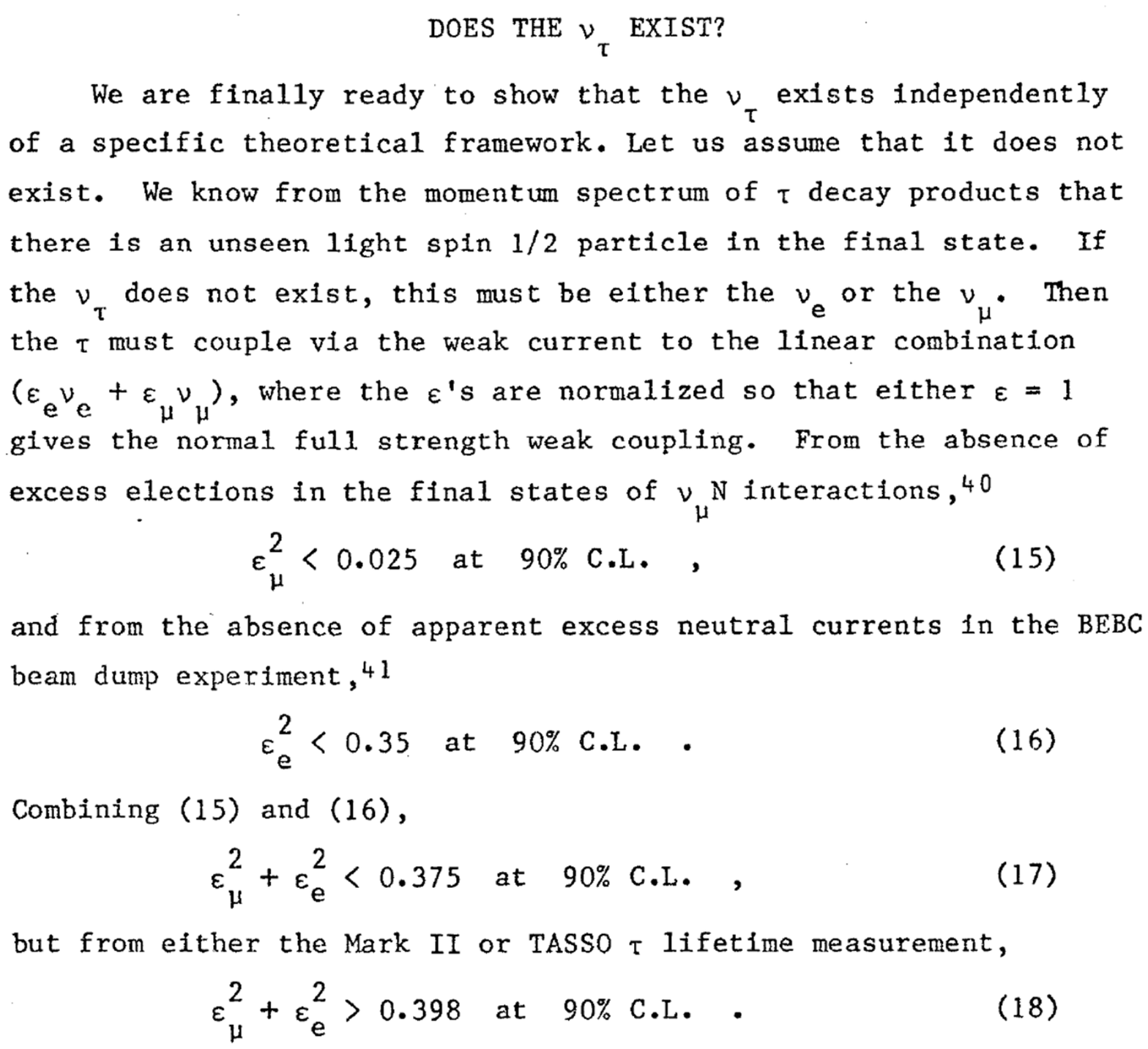}
\vspace{-4cm}
\caption{\label{fig:Feldmans-argument} \small  Reproduction of the section of G. Feldman's conference paper explaining that the existence of the tau neutrino can be demonstrated by confronting the upper limit on the tau life time (i.e. lower limit on the coupling) with the upper limit on the rate of electron and muon neutrino coupling to the tau in neutrino beam experiments}
\end{figure}

The limit for the direct coupling of electron neutrinos to the tau was more difficult to achieve, and would require either a good positive identification of the tau, which would come later, or a higher fraction of electron neutrinos in the beam. The latter was achieved using a beam dump neutrino beam, in which the component of neutrinos from pion and kaon decays is reduced by a factor 1000 to 3000 by using a large copper dump where the interaction length is much shorter than their decay length. Remain neutrinos from charm or b decays. For the CERN beam of 400 GeV protons, b production is still very small and so is the prompt tau neutrino production. Charm itself produces essentially equal rates of muon and electron neutrinos. CERN ran beam dump experiments in 1977-1979, with dumps of various density to make it possible to interpolate to infinite density. The Big European Bubble Chamber, BEBC, filled with a neon-hydrogen mixture of density 0.71 for a radiation length of 44cm, observed a signal of 15-20  prompt electron and muon neutrinos, confirming charm production in hadronic interactions. A tau signal would manifest itself by an increase of events classified as neutral current interactions, because of hadronic tau decays. No such excess was observed and a limit of 0.35 at 90\% C.L for the $\nu_e$ - $\nu_\tau$ mixing was set. The authors also concluded that "the hypothesis that the $\nu_e$ - $\nu_\tau$ are identical
would lead to an anomalously high NC/CC ratio for $\nu_e $  and $ \bar {\nu_e} $.  
Using the value reported above this hypothesis can be excluded at the 90\% confidence level". 

This conclusion did not cover, of course, the more realistic hypothesis described in the early part of this section, of mixing with a heavier state. This would require a measurement of the tau lifetime.

In 1981 Feldman  concluded that the first measurements of the tau lepton lifetime combined  with the absence of tau production in e.g. these neutrino experiments, excluded that the "$\nu_\tau$" could be entirely made of a mix or superposition of 
$\nu_e$  and/or $\nu_\mu $~\cite{Feldman:1981nv}. The argument is reproduced in Fig.\ref{fig:Feldmans-argument}. It followed from this that the  "$\nu_\tau$" could not be entirely composed of the known  $\nu_e$  and/or $\nu_\mu $ and was likely to comprise a new  neutrino. To modern eyes and eighties alike, the statistical significance provided by two 90\% C.L. exclusions  almost touching each-other was not very strong .   

PDG1982~\cite{Roos:1982sd} faithfully reported the milestone achieved by the tau neutrino, by elevating it to the status of "established stable particle", although qualifying the achievement of 'indirect'. Feldman's paper was quoted -- to this day (5 September 2018) it has been cited a total of 5 times. Apparently, the question whether the tau neutrino existed was considered too evident to capture the passion of the masses, and the answer was a bit involved. 
Be as it may, this same wording would remain in the full listing until 2002, although much more significant evidence would be gathered in the following few years. 

\begin{figure}[htb!]
\centering
\vspace{-2cm}
\includegraphics[width=0.95\textwidth]{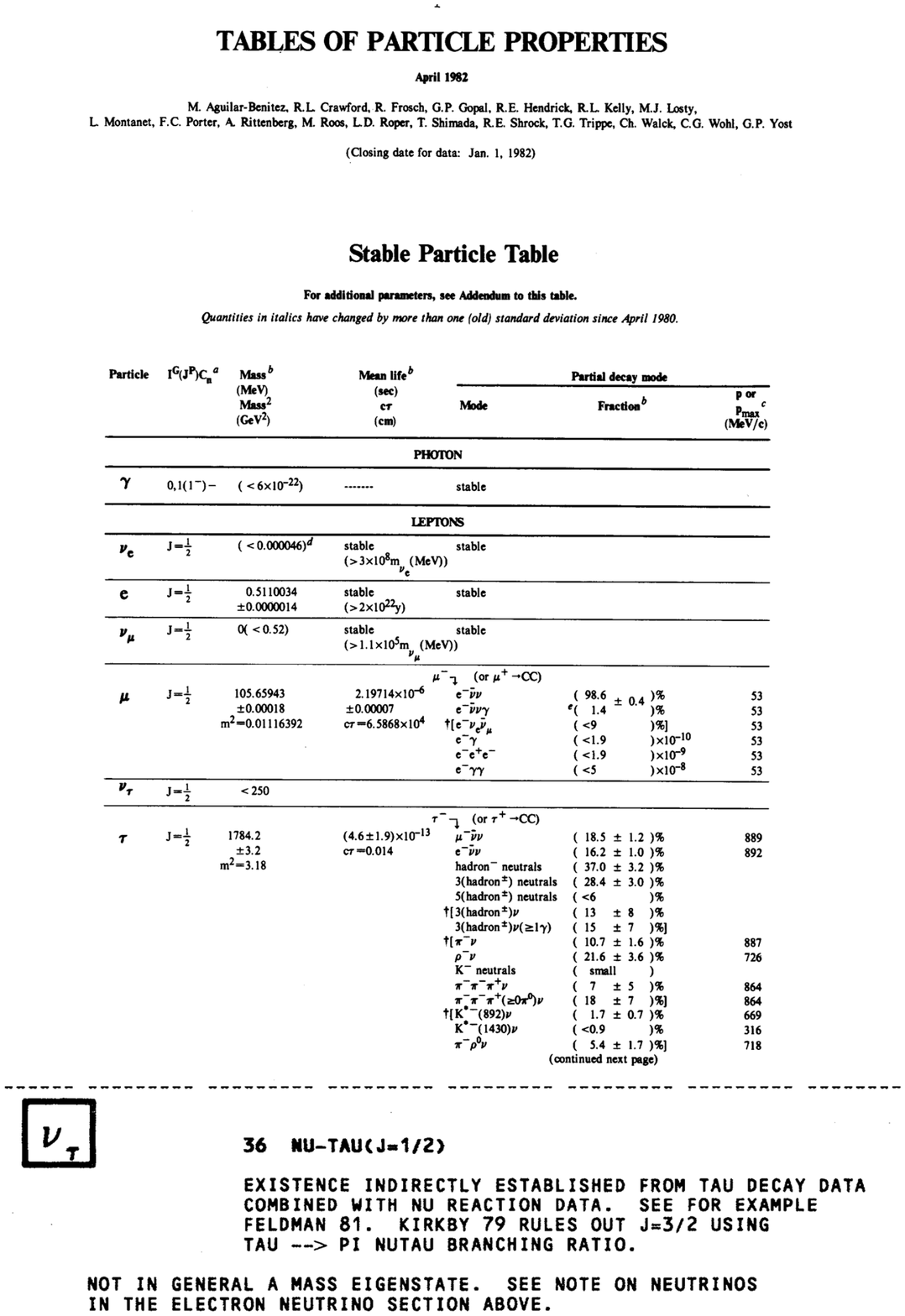}
\vspace{-2cm}
\caption{\label{fig:PDG1982} \small  PDG1982: the title and the Stable Particle Table, showing for the first time the tau neutrino as an established, stable particle, with spin 1/2, m<250 (from $\tau \rightarrow \pi \nu$  decay). The life time measurement is still poor. A large number of hadronic decays is reported; the  $K^*\nu/\rho \nu$ decay ratio is consistent with the Cabibbo angle, a trademark of the hadronic weak decay. At the bottom is reproduced the full listing entry, indicating that the tau neutrino existence is {\sl indirectly} established, referring to Feldman's paper.}
\end{figure}


\newpage
\vfill\eject

\section{The tau neutrino is established}
In the following years the ingredients to the demonstration of existence of the tau neutrino as a particle distinct from the $\nu_e$  and/or $\nu_\mu $ and isospin partner of the tau would be gathered. More precise measurements of the tau lifetime, and a more dedicated search for  neutrino Charged Current events with a tau in the final state were 
made.  

\subsection{improvements in $\tau$ lifetime measurements}
The lifetime measurements of charm, b and tau were of great interest and dedicated vertex chambers were built. This effort was pioneered by the MarkII collaboration with a high resolution drift chamber~\cite {Jaros:1983uq}, soon followed by the TASSO and MAC experiments for an average of $3.3\pm 0.4 ~10^{-13}s$ in the 1986 PDG average~\cite{AguilarBenitez:1986fu}, a factor 7 better than that of 1981. Together with the tau to electron branching ratio of $17.4 \pm 0.5$\%, this established that the tau coupling to the "$\nu_\tau$' was consistant with that of the muon to its neutrino at a level of precision of  $\pm$ 10\%.

\subsection{W decays}
After the discovery in the UA1 and UA2 experiments of the W in the $\rm{W\rightarrow e\nu}$ and  $\rm{ W\rightarrow \mu \nu }$ channels, search continued in particular in the hope of finding the top quark or signals of super-symmetry by means of the missing energy and transverse
momentum signatures. Those of us who were listening to that channel at that time will remember that such missing events were indeed reported, and caused some excitement~\cite{Ellis:1984sz}, until it was realized that what had been observed was the 'Altarelli Cocktail'~\cite{Ellis:1985xw} composed mainly of $\rm{ W\rightarrow \tau \nu }$ and Z decays into neutrinos accompanied with jets. See Di Lella for a recent recollection~\cite{Dilella2016} and this delightful quote: {\sl Experimenters and theorists alike should stop wishful thinking about new physics and [...] start a serious,
quantitative background evaluation}. I let the reader project this moral to his favorite, more recent, situations. 

\begin{figure}[htb!]
\centering
\vspace{-0cm}
\includegraphics[width=0.45\textwidth]{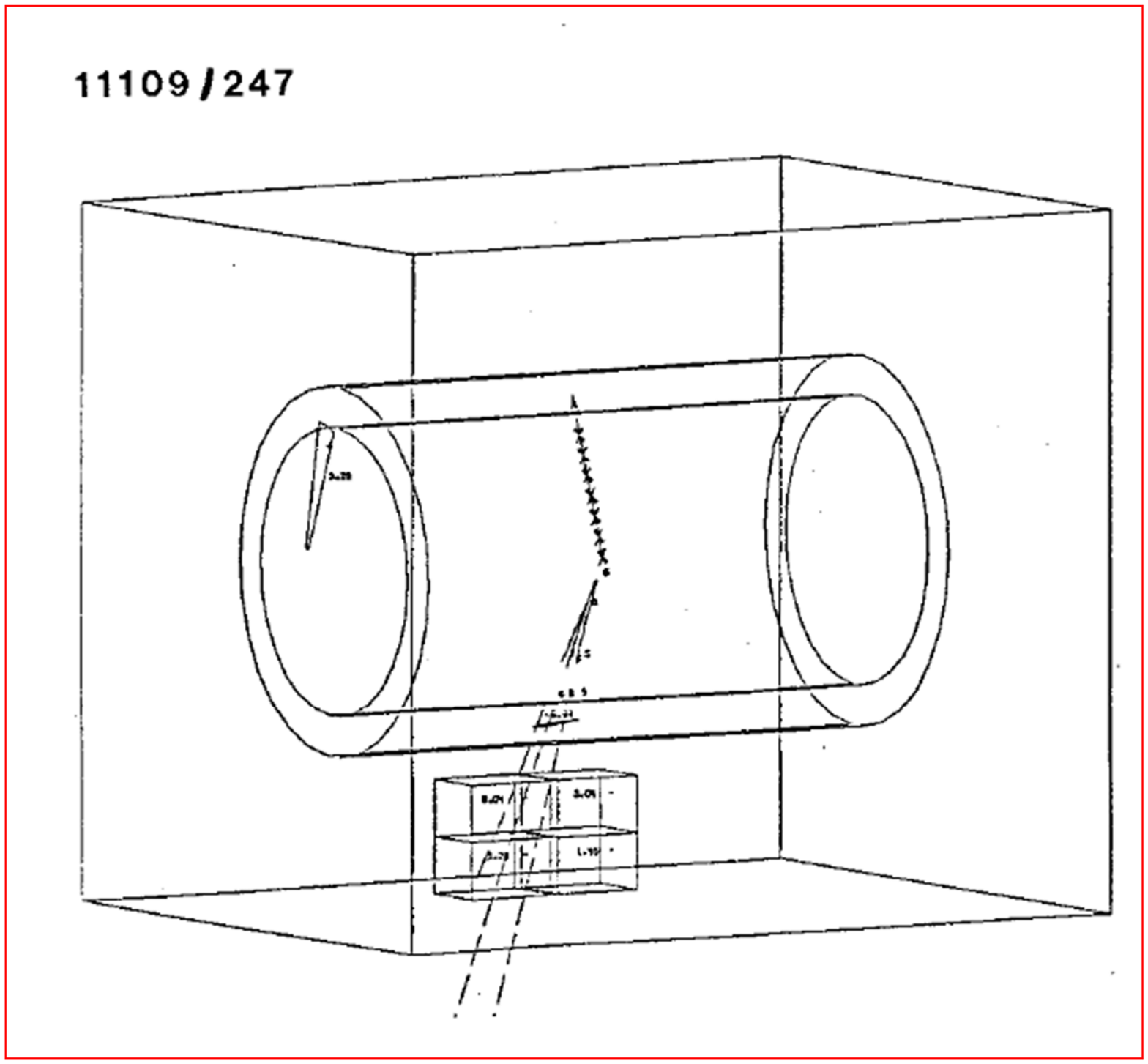}
\includegraphics[width=0.45\textwidth]{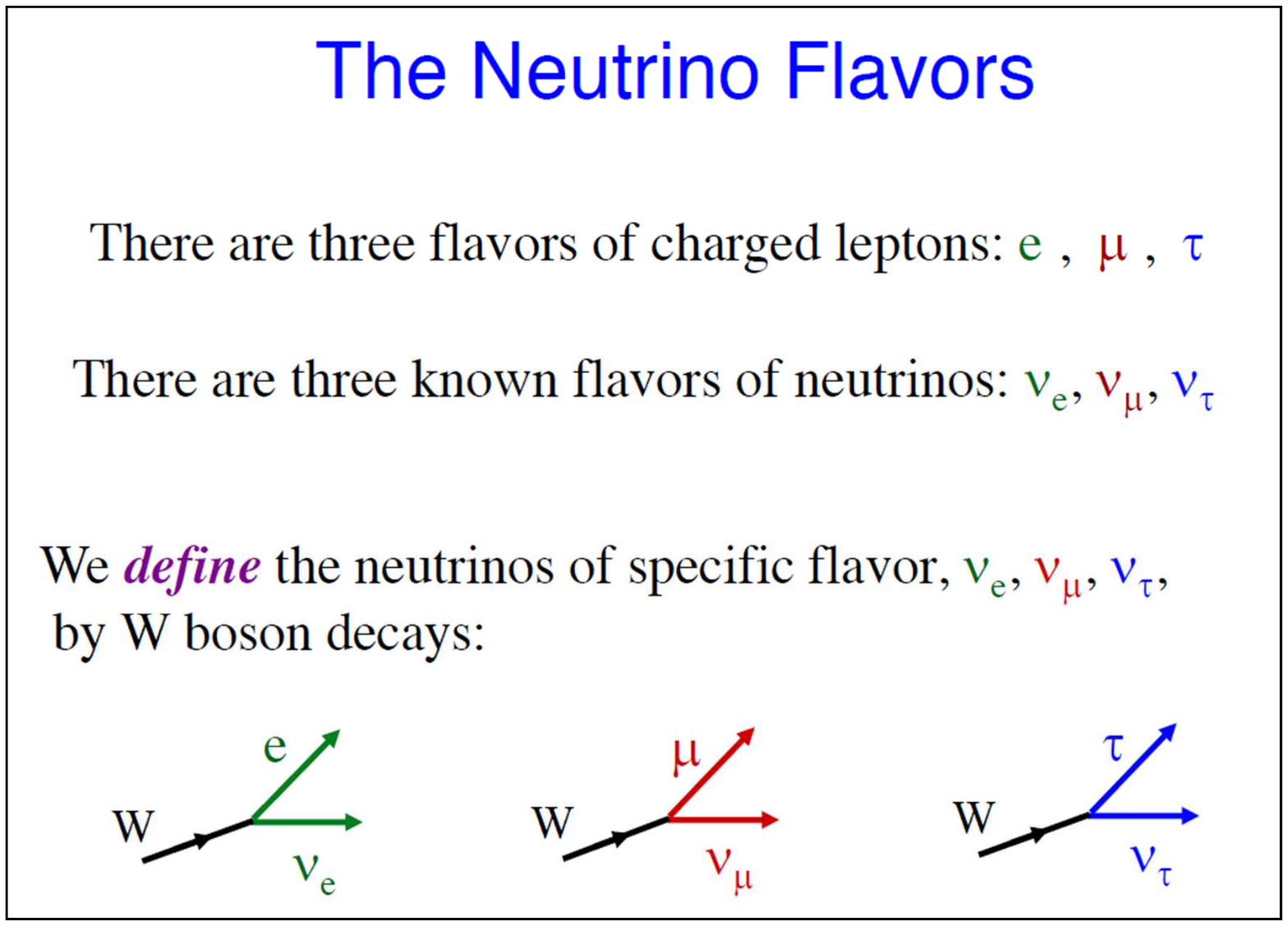}
\vspace{-3cm}
\caption{\label{UA1-W-tau-nu} \small  Left: Observation in UA1 of  the W decay $\rm{W\rightarrow \tau\nu}$ with the tau decaying into 3 pions and a missing transverse momentum of 40 GeV. Right: the definition of neutrino flavour from the W decay (B. Kayser, 
VIIth Pontecorvo School, August 2017)}
\end{figure}

In 1985 the UA1 collaboration reported a handful of $\rm{ W\rightarrow \tau \nu }$ candidates~\cite{Vuillemin:1985tc,Albajar:1986fn}, with a hadronic Jet on one side and missing transverse momentum in the other, as in the example given in Fig.~\ref{UA1-W-tau-nu}. The ratio of decay fractions to taus and electrons  was measured to be $\Gamma_{W\rightarrow \tau \nu}~/~\Gamma_{W\rightarrow e \nu} = 1.01 \pm 0.2 \pm 0.13$, a "direct" measurement of the universality of couplings as predicted in the sequential lepton hypothesis and the Standard Model.  

What appeared at the time like a revenge of the Standard Model, was actually an important moment in the history of the tau neutrino, for its production in (real or virtual) W decay is what is used today as a definition of the neutrino flavour, (Fig.~\ref{UA1-W-tau-nu}, right).

\subsection{E531}
\begin{figure}[htb!]
\centering
\vspace{-0cm}
\includegraphics[width=0.95\textwidth]{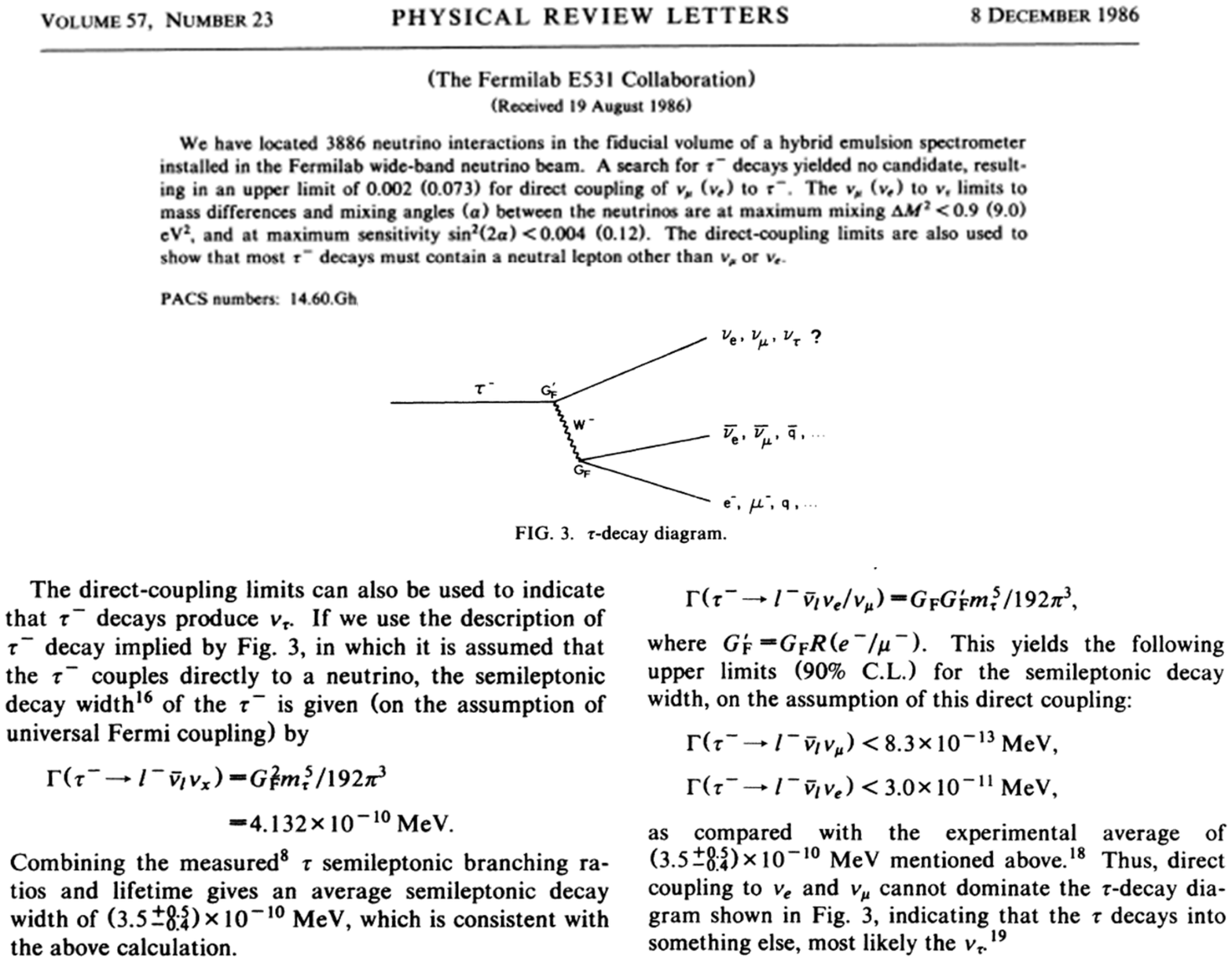}
\vspace{-8cm}
\caption{\label{E531} \small  abstract of the E531 paper and cuttings from the section where the }
\end{figure}

A major improvement came from the Fermilab E531 neutrino experiment~\cite{Ushida:1986zn}. E531 was designed to measure the lifetimes of charmed particles produced by the Fermilab
neutrino beam and produced lifetimes for the 
$ \rm{D^0, D^\pm ,F^\pm, and \Lambda_c^+}$.  This time the tau signature could be "direct", namely  a secondary vertex and no other prompt lepton.  In this way both the electron neutrino and muon neutrinos contained in the beam could be used as normalisation to set a limit. 
Having seen no tau neutrino candidate event out of 1870 CC$_{\nu_\mu}$  events with an identified muon and an estimated
53 CC$_{\nu _e}$ events, the collaboration could proceed put strong limits of 0.002 (0.073) for the direct coupling of the $\nu_\mu $ ($\nu_e$) to the $\tau$, see Fig.\ref{E531}. With the same argument as Feldman's in comparison with the tau lifetime, this paper was able to establish at the ~(approx.) 8~$\sigma$ level that most tau decays must contain a neutral lepton other than $\nu_e$ and $\nu_\mu $.

Thus the neutral spin 1/2 particle observed in tau decays, which has all the properties of a neutrino, interacts with the tau with a strength that is within 10\% of that of the full V-A weak interaction,  which is distinct from $\nu_e$ and $\nu_\mu $ and nearly orthogonal in the quantum sense of it, had been established as the isospin partner of the tau lepton. 

{\bf In 1986, the existence of the tau neutrino was solidly established}. 

\subsection{Discussion}
\label{Discussion}

I had not been personally involved in any of the papers that are quoted earlier although I did sign a few of the MarkII papers. I got interested in the question, and had several discussions with Jack Steiberger, Friedrich Dydak and Tony Pich, while preparing actively in 1988-1989, as ALEPH physics workshop convener,  to execute the measurement of the number of neutrino families -- would we discover the tau neutrino? Having recently moved offices (Aug. 2018) I retrieved a few papers that I had been reading at that time; below are a few quotes -- all calling for a 'direct' experiment to observe tau neutrino interactions.    
\begin{itemize}
\item B. Barish and R. Stroyonowski~\cite{Barish:1987nj} 
{\sl Measurements of the lifetime of the $\tau$ plus these (reference to E531) limits provide indirect evidence that the  $\nu_\tau$ is distinct from $\nu_e$ and $\nu_\mu $. On the other hand, there have been no
direct observations of $\nu_\tau$, that is, interactions resulting from neutrinos produced in tau lepton decay}. 

\item Friedrich Dydak~\cite{Dydak:171238} 
{\sl However the $\nu_\tau$ remains today's best established yet unobserved particle...} 

\item Antonio Pich~\cite{Pich:1989qk} 
(After a couple paragraphs summarizing the knowledge of the existence of the tau neutrino as distinct from $\nu_e$ and $\nu_\mu $, including quote to E531). {\sl However, no direct
observation of $\nu_\tau$, that is, interactions resulting from neutrinos produced in tau
decay, has been done so far.} 

\item Denegri, Sadoulet and Spiro~\cite{Denegri:1989if}
{\sl Until now the direct detection of neutrinos has been achieved only for $\nu_e$ and $\nu_\mu $. The third generation $\nu_\tau$ has not yet been directly observed through its characteristic interactions with matter. The evidence for  $\nu_\tau$ as an independent species, with (universal)  Fermi coupling (...), is indirect. It is obtained from the tau lifetime (references) or (...) from the W partial production cross-section ratios (...) (reference to UA1).  }

\item K. Winter~\cite{Winter:1989mj} 
{\sl Does the tau-neutrino exist as a particle? Surprisingly, this question cannot be answered by yes or no. Its existence can be proved by direct observation of the charged current reaction ${\rm \nu_\tau +N \rightarrow \tau^- +X}$.} 
(NB this paper was originally presented at a conference in Russia in March 1989 and published in acta physica Hungarica 68 (1-2) pp. 135-143.) 
\end{itemize}

So there seems to have been a collective feeling of dissatisfaction at the way the tau neutrino had been discovered, and the desire for a 'direct' observation of tau neutrino interactions with matter was widely expressed. It is difficult to trace the origin of this desire, except for the impression that the tau neutrino should be properly discovered like the other neutrinos by their interaction with matter.

One thing is certain: none of these calls was backed up by a reference that would explain what, from this observation, would be learned that was not already known.  Also, most of these papers refer to the E531 paper, although not necessarily in the quote above. 

In all the arguments above two primary facts are missing: 
\begin{enumerate}
\item the tau had actually been {\sl observed} to decay into a neutrino, for which the question was not of its existence but of its nature; nature which had taken until 1986 to firmly establish, by a combination of several experimental facts. 
\item The tau having been shown factually to be decaying into some neutrino state "$\nu_\tau$", it is clear that if one could make a beam of "$\nu_\tau$", they would NECESSARILY interact giving taus with a rate related to the tau lifetime, irrespective of the nature of "$\nu_\tau$", for instance whether it is a coherent superposition or mixing of $\nu_e$ and $\nu_\mu $  or pure $\nu_\tau$. 
\end{enumerate}

So why was the existence of the tau neutrino declared 'established indirectly'? My first reactions was to assimilate 'direct' with a mountain itinerary. Direct  would be that which was simplest, indirect would be a more complicated route -- though maybe easier. In scientific term, maybe direct would mean 'can be understood using zero or one equation'. Then 'indirect'  proof would require two or more equations. In that case there would be no scientific hierarchy between direct and indirect and an indirect proof would be just as good. 

There seemed to be more to it. After giving it a bit of thought I concluded that 'indirectly' could in this case only be {\sl defined} by contrast with the 'direct' "observation of $\nu_\tau$, {\bf that is}, (i.e. {\sl by definition}), interactions resulting from neutrinos produced in tau lepton decay". In this case like in many others the word 'direct' has a context-sensitive definition, and 'indirect' being its opposite, is even worse.    

We had several discussions on this topic with Jack Steinberger before and after he received his Nobel Prize in fall 1988. He consistently considered that the existence of the tau neutrino having already been established, (for which he referred to 'that Fermilab experiment') we couldn't discover it a second time. More later.

\section{What is the number of families of neutrinos?}
\label{sec:introduction}

At the time SLC and LEP started, the basic properties of the weak interaction
were already well known.  
One pressing question, however, could not be answered either by 
theoretical arguments or by direct experiments: 
what is the number of families of fermions?
LEP answered this fundamental question, at least for the light active neutrinos, in a few weeks 
by measuring the Z resonance.   
With six years of data and meticulous measurements 
of luminosity and energy  
the LEP experimentalists determined the Z boson mass and 
width, as well as the Z decay rates, 
with a  precision which is unlikely to be surpassed soon.

All elementary quarks and leptons that have been observed 
are organised in {\em exactly} three families: 
(or generations).    
\[
\begin{array}{c c c c l}
\left(\begin{array}{c} u\\d' \end{array}\right)&
\left(\begin{array}{c} c\\s' \end{array}\right)&
\left(\begin{array}{c} t\\b' \end{array}\right)&&
\mathrm{doublets~of~left-handed~quarks}\\
&&&\\
\begin{array}{c} (u)\\(d) \end{array}&
\begin{array}{c} (c)\\(s) \end{array}&
\begin{array}{c} (t)\\(b) \end{array}&&
\mathrm{singlets~of~right-handed~quarks}\\
&&&\\
\left(\begin{array}{c} \nu_e\\ e       \end{array}\right)&
\left(\begin{array}{c} \nu_\mu\\ \mu   \end{array}\right)&
\left(\begin{array}{c} \nu_\tau\\ \tau \end{array}\right)&&
\mathrm{doublets~of~left-handed~leptons}\\
&&&\\
\begin{array}{c} (?\nu_e)\\(e)        \end{array}&
\begin{array}{c} (?\nu_\mu)\\(\mu)    \end{array}&
\begin{array}{c} (?\nu_\tau)\\(\tau)  \end{array}&&
\mathrm{singlets~of~right-handed~leptons}
\end{array}
\]

As far as we can tell the Electroweak theory could 
not easily accommodate further isolated
fermions, but it could accommodate any number of families of the same type.  
One could easily envisage a situation where many families
including heavy charged quarks and leptons could exist, 
without these heavy leptons being ever produced in
accessible experiments, by lack of available energy. 
Nevertheless, since the known
neutrinos are very light,
it is natural to expect that these additional 
families would include light neutrinos as well,
leading to the possibility that many families of light neutrinos would exist. 

The existence of many light neutrinos would have considerable 
cosmological consequences. 
In particular, the evolution of the universe within 
the first second after the Big Bang 
would be profoundly affected. The argument, developed in~\cite{Weinberg:1972kfs}, 
is the following.
At the time where energies are large enough, reactions such 
as $e^+ e^- \rightarrow \nu\bar{\nu}$
transform a fraction of available energy into neutrinos in a democratic way. 
The creation of 
neutrons and protons however is controlled by reactions involving 
the electron neutrino, such as 
$\nu_e + n \rightarrow p + e^-$, and is consequently very sensitive 
to the number of light
neutrino families, $\nnu$, which compete with electron neutrinos.
The relative abundance of Hydrogen, 
Deuterium and Helium, 
and therefore the entire 
chemical constitution of our universe is a sensitive function of this number.  

Before SLC and LEP started, limits on the number of light neutrinos were 
given from the above cosmological considerations, since there are data on the 
relative abundance of various nuclei in the universe, in particular the ratio
 of helium to hydrogen, or, with similar arguments, 
 from the time development of the supernova 1987A. There were also indications 
from the direct search for the process
 $\epem \rightarrow \nu \overline{\nu} \gamma$ (single-photon experiments), or from the
early measurements of the Z and W boson properties in the CERN and FERMILAB
${\rm p \overline{p}}$ experiments. 
A review of these constraints published 
in 1989~\cite{Denegri:1989if} evaluated the best estimate 
of $\nnu$ to be $\nnu = 2.1^{+0.6}_{-0.4}$, and stated that
  ''$\nnu=3$ is perfectly compatible with all data, but four families still 
provide a reasonable fit''.

In searching for further families of neutrinos, 
it will be assumed that their couplings are the same as 
those of $\nu_e$, $\nu_\mu$ and $\nu_\tau$.
Universality is deeply embedded in the Standard Model, 
identical multiplets having the same 
coupling constant. It is very well verified for 
Charged Current interactions of e-$\mu$-$\tau$ 
leptons, including the neutrinos, 
and for Neutral Current interactions
of charged leptons.

\section{Determination of the 
number of light neutrino species at LEP and SLC}
\label{sec:line-shape}

\begin{figure}[htb]
\vspace*{-.5cm}
\centerline{\includegraphics[width=0.8\textwidth]{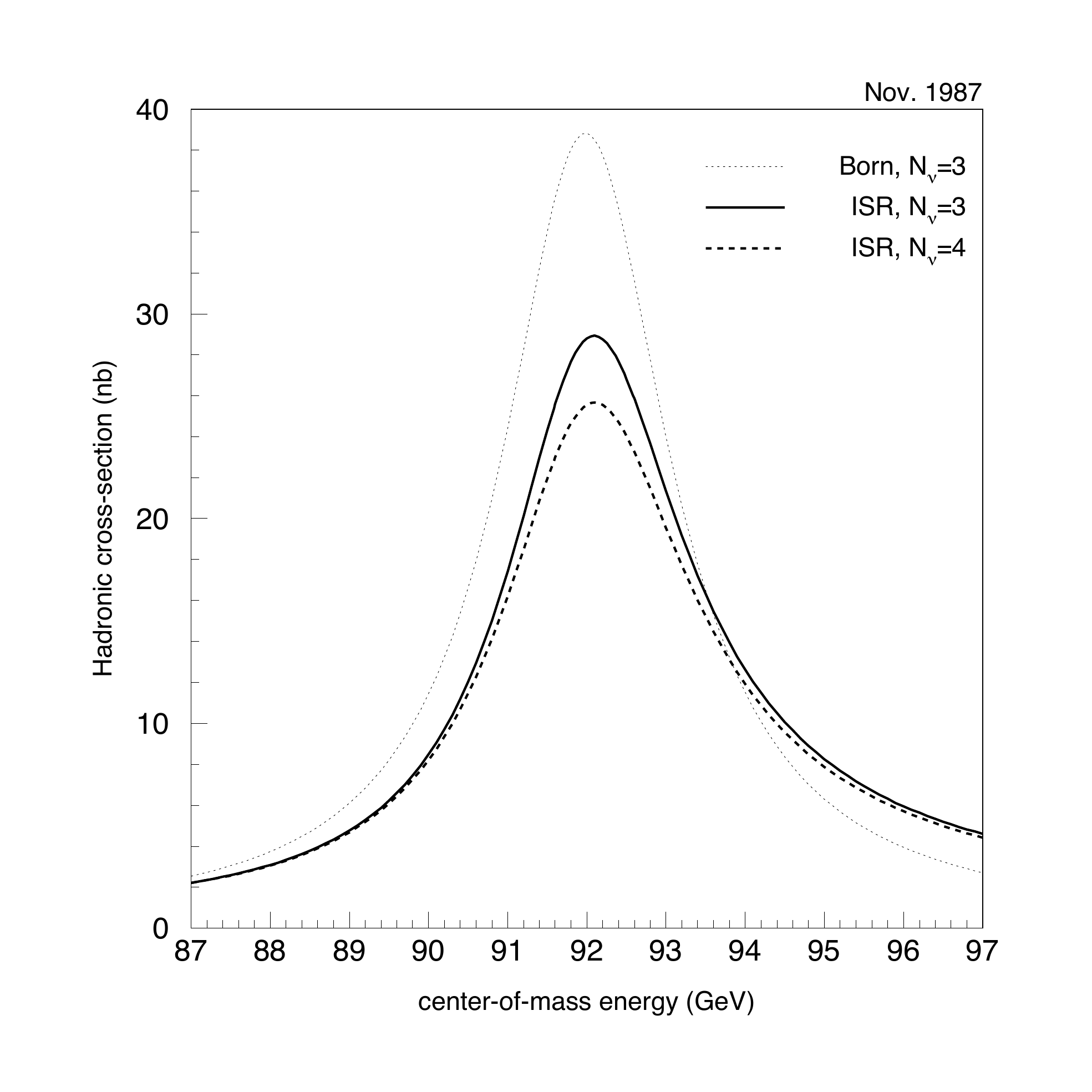}}
\vspace*{-0.5cm}
\caption[The 1987 hadronic line shape]
{{\sl
The $\epem \rightarrow {\mathrm hadrons}$ cross section
as a function of center-of-mass energy. This curve was drawn in 1987, before 
LEP start-up. At that time the Z mass was measured to be around 92 GeV
with an error of 1.5 GeV. 
The doted line represents the born approximation prediction 
for three species of  light neutrinos.
The full line includes the effect of initial state radiation.
The dashed line represents the effect of adding one more type of light neutrino
with the same couplings as the first three. It is clear from this
picture that the cross-section at the peak of the resonance
contains most of the information on the number of light neutrino species.
}}
\label{Zline}
\end{figure}

The most precise determination of the number of light neutrino species 
is obtained from measurements of the visible 
cross-sections of $\epem$ annihilation 
at and around the Z resonance, as is made explicit in
figure~\ref{Zline}. If the Z is allowed to decay into more types of
light neutrinos which will lead to an invisible final state, 
it will decay less often into the visible ones.  
  
The realization that the visible cross-sections 
might be sensitive to the number of 
light neutrinos is rather ancient, one finds the question asked in John Ellis
 ``Zedology''~\cite{Ellis:875312}:
{\em The Z peak is large and dramatic, as long as there are not too many 
generations 
of fermions. Is it conceivable that there might be so many generations as 
to wash out the Z peak?}. 
Since at that time the bound on the number of light neutrinos 
was very weak (about 6000), this certainly was a frightening possibility
for those planning to build LEP! 
Dramatic also were the few first 
weeks of SLC and LEP operation where 
it was quickly realized that the Z peak was there indeed, 
large and dramatic, and that, alas, the number of light neutrinos was three. 

\begin{figure}[htb]
\centering
\vspace*{-2cm}
\includegraphics[width=0.8\textwidth]{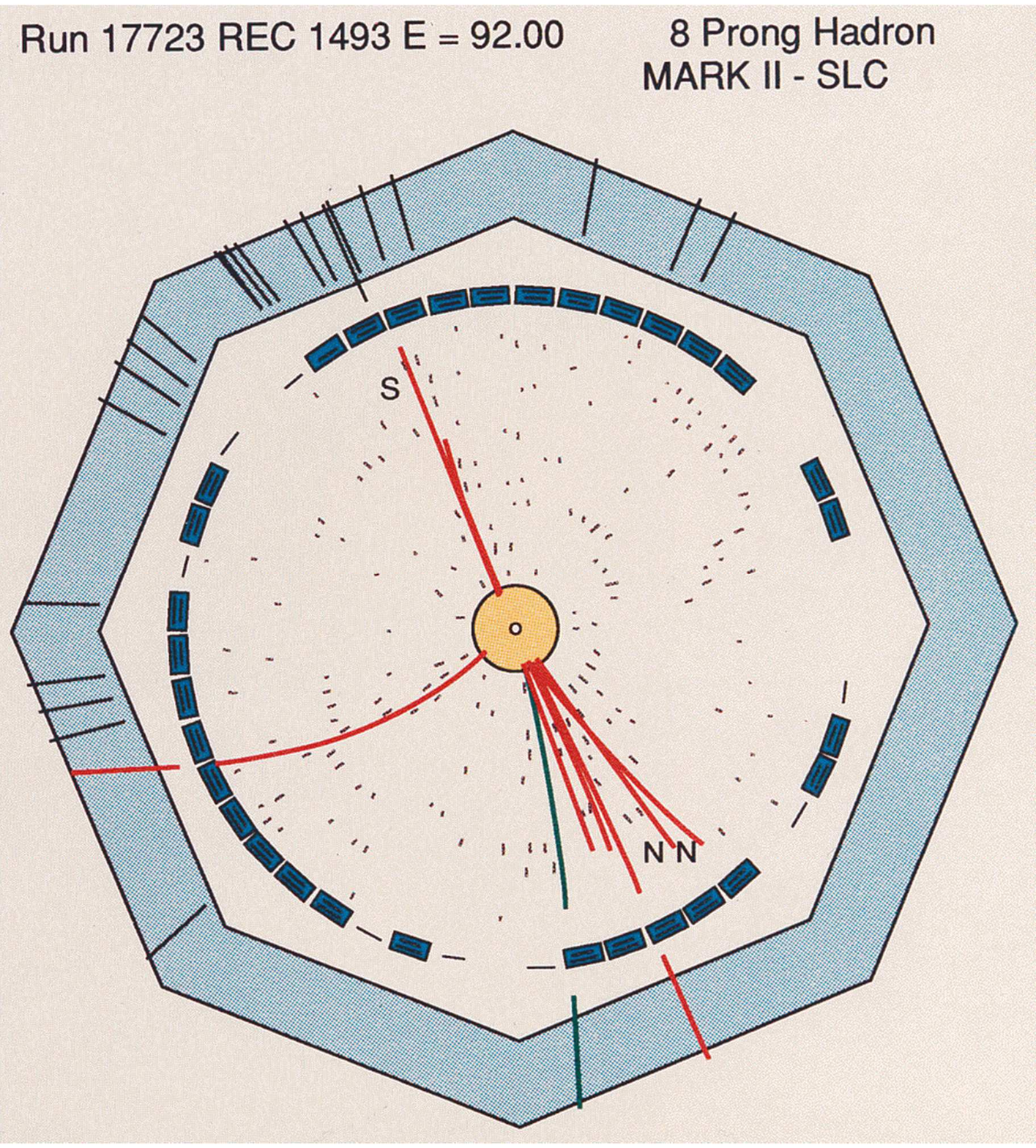}
\vspace*{-2cm}
\caption[The First Observed Hadronic Z Decay]
{{\sl
The First hadronic Z decay recorded in the MARKII detector at the SLC on 12 April 1989
}}
\label{First-Z-MarkII}
\end{figure}

There was intense competition between the SLC at SLAC (California, USA)  
and LEP at CERN (Switzerland).
The two projects were with rather different in concept. 
LEP was build as the largest possible conventional 
$\epem$ storage ring, with a circumference of 27 Km. 
This standard technique would ensure few surprises and  
and reliable high luminosity.
SLC on the other hand, was the prototype of a new concept of accelerator,
the linear collider; it was re-using the old Stanford linac, 
with improvements in the acceleration technique (SLED) 
and addition of arcs to bring $e^+$ and $e^-$ in collisions, as well
as of challenging positron source and damping rings.

The commissioning of SLC started in early 1987, 
and lead to a number of technical difficulties, 
not surprising in retrospect for such a new project.
The first Z hadronic decay  was produced on April 11 1989, and recorded 
in the MarkII detector~\ref{First-Z-MarkII}. Luminosity was very low, a few 10$^{-27}$/cm$^2$/s,
leading to a few Z hadronic decays per day.
With the LEP start-up advertised for the 14 of July, the time where SLC would 
hold the lead was going to be short, and intense.
Nevertheless the SLC collaboration was able to 
collect a total of 106 Z decays by 24 July and submit 
a publication~\cite{Abrams:1989aw}, where the Z mass was determined to be 
$\mz=91.11\pm0.23$~GeV, and the number of light neutrinos species
$\nnu=3.8\pm1.4$.

\begin{figure}[htb]
\centering
\vspace*{-2cm}
\includegraphics[width=0.8\textwidth]{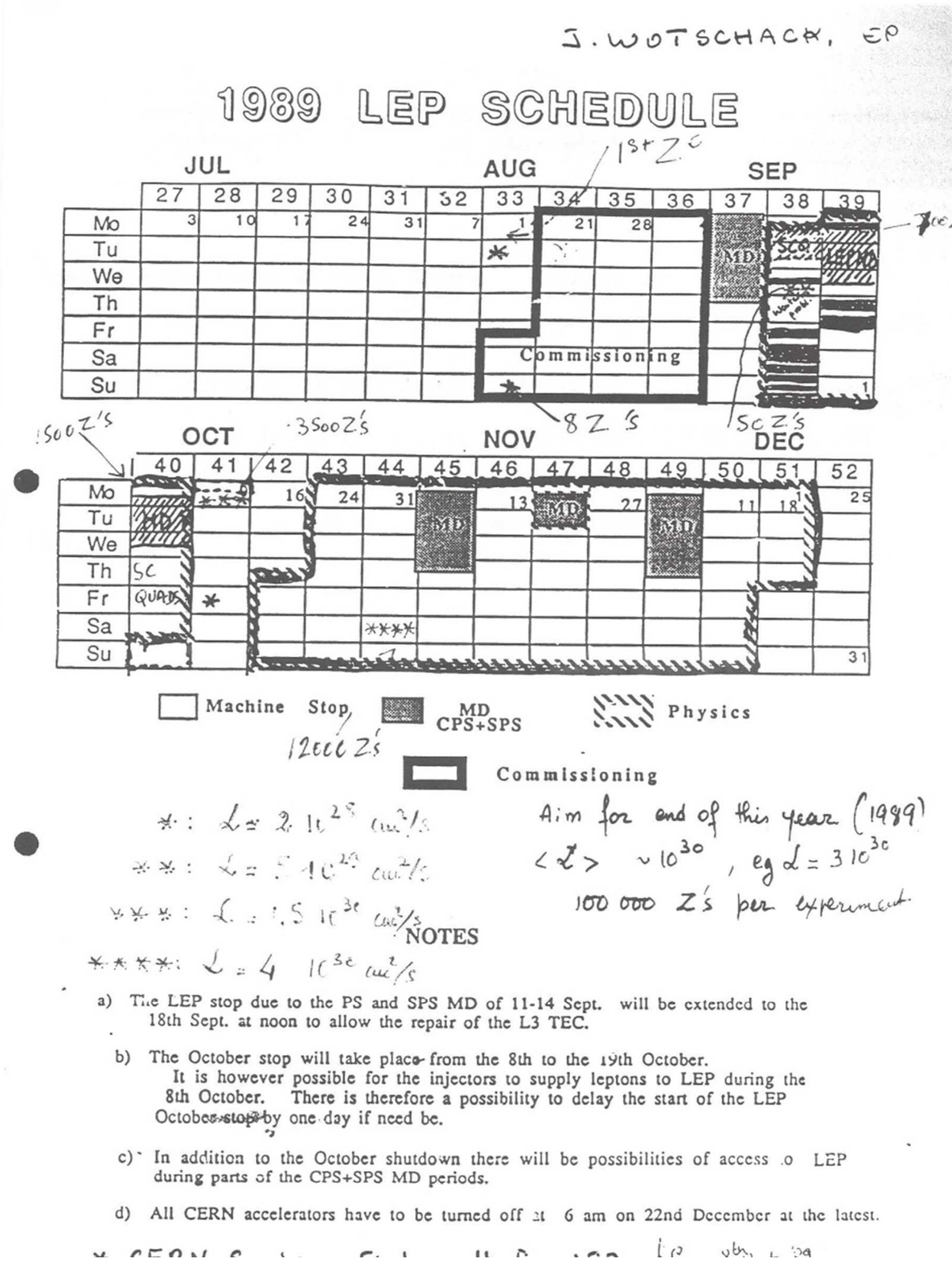}
\vspace*{0cm}
\caption[The 1989 LEP schedule as shown to ALEPH collaboration]
{{\sl
The 1989 LEP schedule as shown to ALEPH collaboration by Jorg Wottschack. The Z event numbers recorded in ALEPH are shown, starting on 13 August 1989.  
}}
\label{LEP-schedule-1989}
\end{figure}

\begin{table}[htb]\centering

\caption{ First results from LEP and SLC on the 
Z mass and the number of light neutrino species, as published around 
12 October 1989 (in order of submission to the journal).} 
{\begin{tabular}{|l| r | r r r | r r r |}
\hline
Experiment & hadronic Zs 
& \multicolumn{3}{c|}{Z mass (GeV)}&\multicolumn{3}{c|}{$\nnu$}\\
\hline
MARKII & 450 & 91.14 & $\pm$ & 0.12 & 2.8 &  $\pm$ & 0.60 \\
L3 & 2538 & 91.13 & $\pm$ & 0.06 & 3.42 &  $\pm$ & 0.48 \\
ALEPH & 3112 & 91.17 & $\pm$ & 0.05 & 3.27 &  $\pm$ & 0.30 \\
OPAL & 4350 & 91.01 & $\pm$ & 0.05 & 3.10 &  $\pm$ & 0.40 \\
DELPHI & 1066  & 91.06 & $\pm$ & 0.05 & 2.4  &  $\pm$ & 0.64 \\
\hline
Average &  & 91.10 & $\pm$ & 0.05 & 3.12 &  $\pm$ & 0.19 \\
\hline
\end{tabular} }

\label{results1989}
\end{table}
 
\begin{figure}[htb]
\centering
\vspace*{-2cm}
\includegraphics[width=0.8\textwidth]{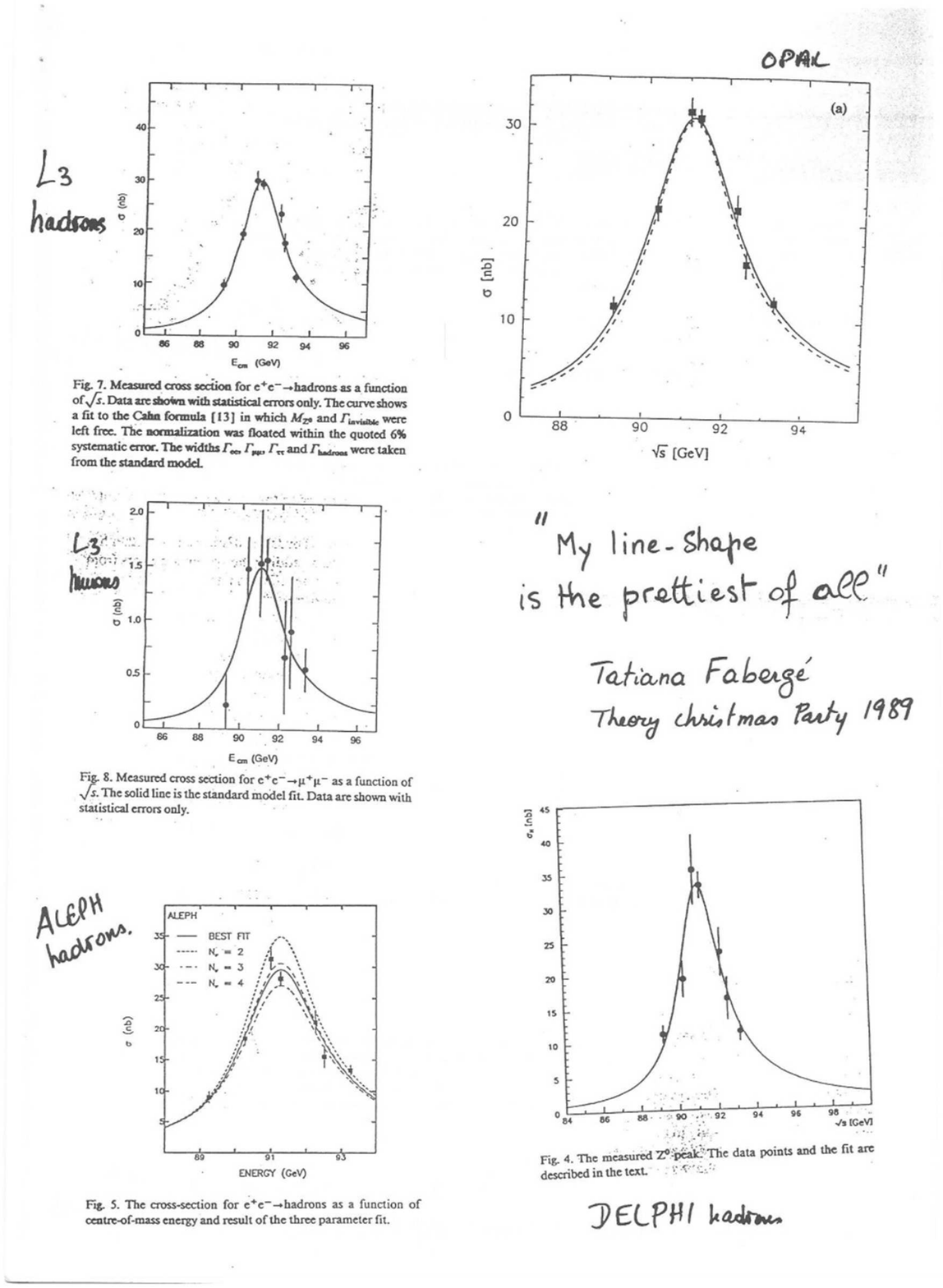}
\vspace*{0cm}
\caption[The first LEP line shapes]
{{\sl
The 1989 LEP line shapes as presented in the seminar on 13 October 1989. A comment extracted from the 1989 TH Christmas party is also reported (from lectures given in Cargèse 1990).   
}}
\label{Line-Shapes-LEP}
\end{figure}

LEP had first beams on 14 of July and collisions on 8 August for one week. The LEP 1989 schedule is shown on Fig.~\ref{LEP-schedule-1989}. 
The high luminosity optics were however not yet commissioned and 
events came at a rate of 1 a day for the four experiments; 
this was not enough to make a measurement. 
Running resumed on  20 September with superconducting
quadrupoles and  in just three weeks, until 9 October,  
around 3000 Zs were collected in each of the experiments.
By 13 October, a seminar was organized at CERN where 
the four collaborations presented their first 
results: L3~\cite{Adeva:1989mn}, ALEPH~\cite{Decamp:1989tu}, OPAL~\cite{Akrawy:1989pi}and DELPHI~\cite{Aarnio:1989tv};  
the results are compiled in Table~\ref{results1989}.
The day before, SLC had organized a public conference where updated
MarkII ~\cite{Abrams:1989yk} competitive results had been presented. 
The ``online average'' of these results is also shown in 
Table~\ref{results1989},  $ \nnu = 3.12   \pm  0.18$. 
The number of light neutrinos was three.

Following this important contribution, 
SLC was shaken by an earthquake on 24 October 1989,
from which it took more than a year to recover, 
and then concentrated on polarized beam physics. 
LEP went on, to the end of  1989 and for 6 more   
years (1989 to 1995), each experiment collecting 4 million hadronic Z decays.
With the final results now available, the 
number of light neutrinos determined to be   
$ \nnu = 2.9841  \pm  0.0082$~\cite{ALEPH:2005ab}.

The early results were unexpectedly precise,
and the precision of final ones exceed by a factor 100 
the expectations that could be found 
in the studies preceding the start of LEP.  
Once the method is explained  in more detail, 
it will become clear that the unexpected 
capacity of the experiments to perform precise 
measurement of hadronic cross-sections
is the reason for this success. 

To finish this section let me quote the conclusion of the ALEPH paper~\cite{Decamp:1989tu}: "The demonstration that there is a third neutrino confirms that the $\tau$ neutrino is distinct from the $e$ and $\mu$ neutrinos. The absence of a fourth light neutrino indicates that the quark-lepton families are closed with the three which are already known except for the possibility that higher order families have neutrinos with mass in excess of ~30 GeV." The first sentence had been written by Jack Steinberger; I wrote the second one. This latter statement is now also demonstrated for quarks, the fact that the Higgs boson at the LHC is produced by gluon-gluon fusion at the expected cross-section ruling out that further families of quarks heavier than the top contribute to this loop diagram.   

\newpage
\vfill\eject

\section{Determination of the Z line shape parameters}
\label{subsec:method}
  
Around the Z pole, the $e^+e^- \rightarrow Z \rightarrow f\bar{f}$
 annihilation cross-section is 
given by 
\begin{equation}
\sigma_f = \frac{12\pi(\hbar c)^2}{\MMZ} \frac{s\Gamma_e \Gamma_f
}{(s-\MMZ)^2+s^2\frac{\Gz^2}{\MMZ}}.
\label{sigmaf}
\end{equation} 
which is a general formula or a spin one particle produced in  
$\epem$ annihilation into a visible channel $f$. This  typical
resonance shape peaks around the Z mass, $\sqrt{s}=\mz$, and
has a width $\Gz$. If the Z decays a fraction $B_f$ of the time into a
final state $f$ the corresponding partial width is defined as $\Gamma_f = B_f \Gz$.

The Standard Model predicts the  numerical values for the Z partial widths, as displayed in table~\ref{couplings}. 
The main decay mode of the Z is into hadrons (70\%), 
each of the leptons representing only $3\%$ and three neutrinos would 
contribute $20\%$. There are no other substantial decay modes unless there
are new particles; the Higgs branching ratio, in particular, 
is very small.  
In this expression the number of neutrinos intervenes 
through the total Z width $\Gz$:

\begin{equation}
\Gz = 3 \gamlep + \GH + \nnu \gamnu
\label{gamz}
\end{equation}  

If $\nnu$ increases, the total width which is in the denominator of 
equation~\ref{sigmaf} 
increases, and the cross-section is 
decreased.

\begin{table}[htb]\centering
\caption{}{Numerical values of quantum numbers, Neutral Current couplings,
and Z decay partial widths,
for the four types of fermions, for hadrons and total width. 
The value of $\sef$ is 0.2315}
{\begin{tabular}{@{}|c|c c|c c|c |@{}} 
\hline

$f$          & $I_{3f}$  &  $Q_f$  &  $\sgaf$  &  $\sgvf$  &  $\Gamma_f$ (MeV)    \\
\hline
$\nu$        &   1/2     &   0     &   1/2   &   1/2   &   167    \\
$e$        &  -1/2     &   -1    &  -1/2   &  -0.04  &   84\\
$u$          &   1/2     &   2/3   &   1/2   &   0.19  & 300  \\
$d$          &  -1/2     &  -1/3   &  -1/2   &  -0.35  & 383\\
$b$          &  -1/2     &  -1/3   &  -1/2   &  -0.35  & 376\\
${\mathrm{hadrons}}=u + d + c +s +b$   &  --     &  --   &  --   &  -- & 1740\\
${\mathrm{total~for~ three~ neutrinos}}
$         &  --     &  --   &  --   &  --  & 2500\\
${\mathrm{total~ for~ four~ neutrinos}}
$         &  --     &  --   &  --   &  --  & 2670\\
\hline
\end{tabular}}
\label{couplings}
\end{table}

The equation~\ref{sigmaf} receives a number of modifications to account for 
the contribution of the photon exchange process (this is less than one percent),
 and more importantly  what is called 'initial state radiation' (ISR), 
in which one or both of the initial state electrons lose energy into photons. 
This phenomenon reduces the initial state energy and smears out the resonance
significantly as can be seen in figure~\ref{Zline}. Due to the availability
of calculations up to second order in perturbation theory, 
this large (30\% at the peak) 
effect can be corrected for with a relative precision of $5 10^{-4}$.

\begin{figure}[h]
\vspace*{-.5cm}
\rotatebox{-90}{\includegraphics[width=0.95\textwidth]{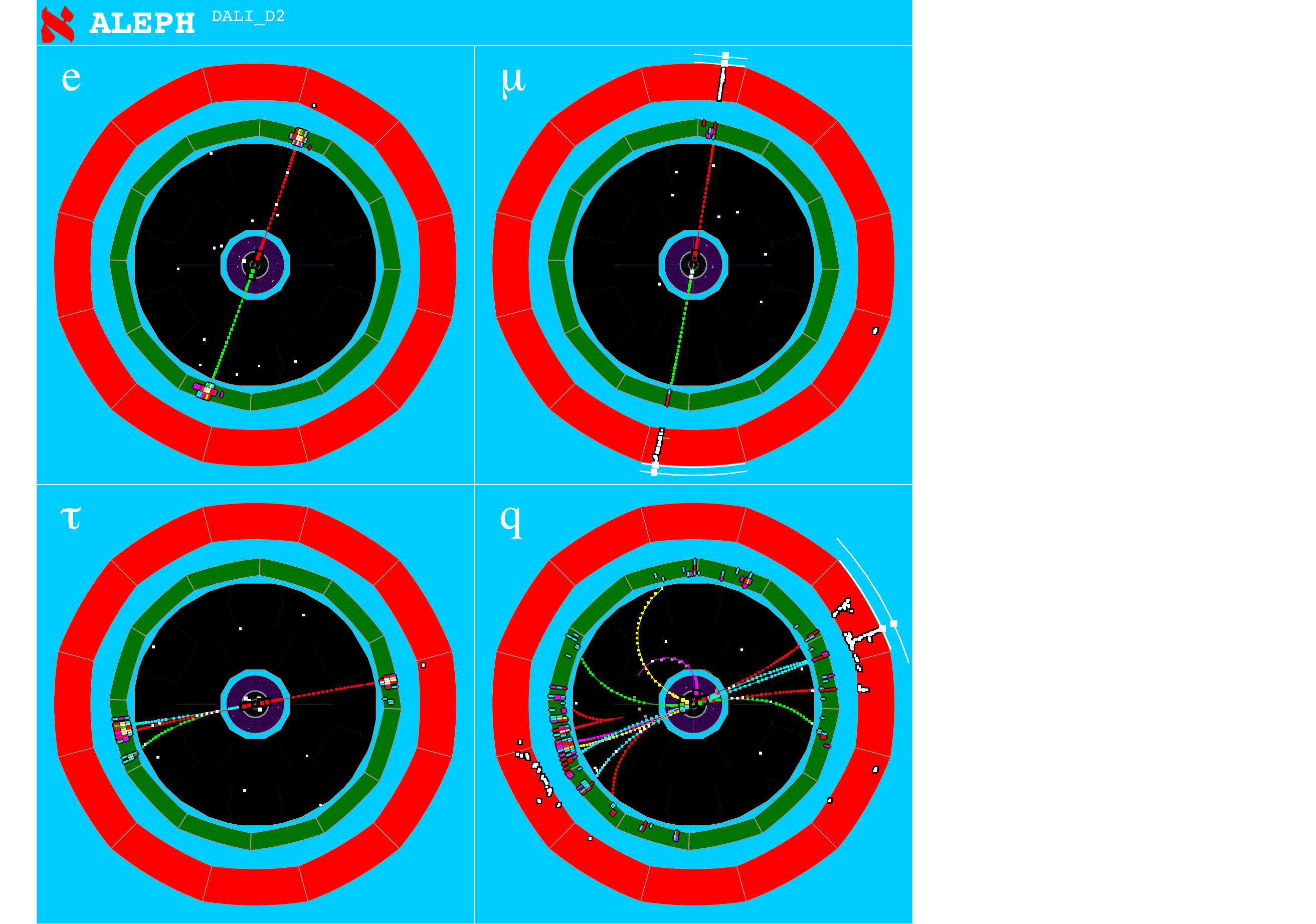}}
\vspace*{-0.5cm}
\caption[Z boson decays]
{{\sl
The four types of Z boson decays. Up left a decay into a pair of electrons,
 up right into a pair of muons, bottom left a pair of tau leptons, one
decaying into an electron and the other into three particles, 
bottom right a pair of quarks that fragments into a number of hadrons.
}}
\label{Zdecays}
\end{figure}

The principle of the analysis is then as follows:
all visible channels are detected by large acceptance detectors
and classified according to four categories: 
i)~hadrons, ii)~electron pairs, iii)~muon pairs, iv)~tau pairs. Examples of 
such events are shown in figure~\ref{Zdecays}. These events are easy to detect,
with high and well known efficiencies (as high as $99 \pm 0.05\%$ 
for hadronic decays), and to separate from each other.

\begin{figure}[h]
\vspace*{-.5cm}

\parbox{0.34\textwidth}{\includegraphics[width=0.30\textwidth]{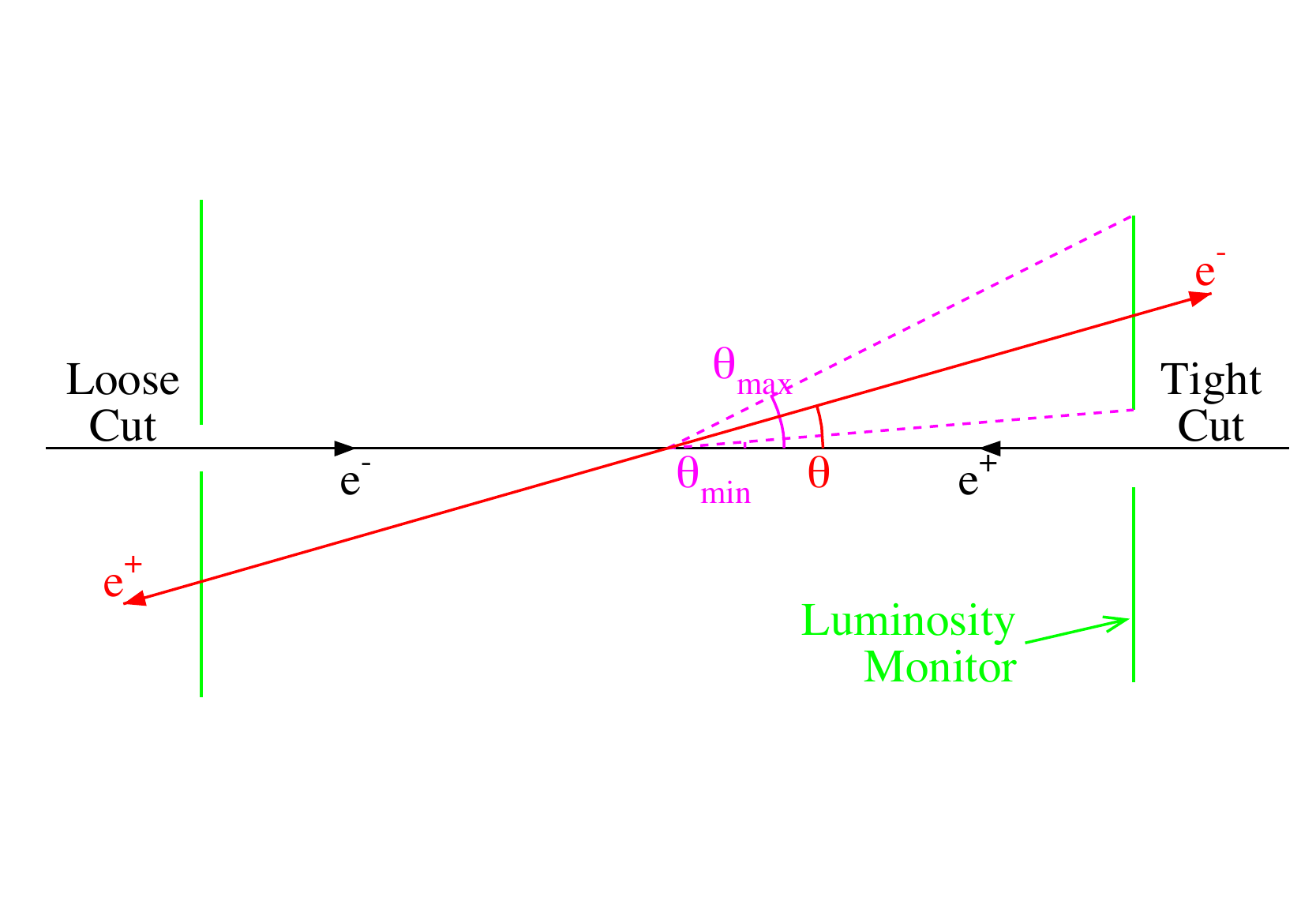}}
\parbox{0.64\textwidth}{\includegraphics[width=0.6\textwidth]{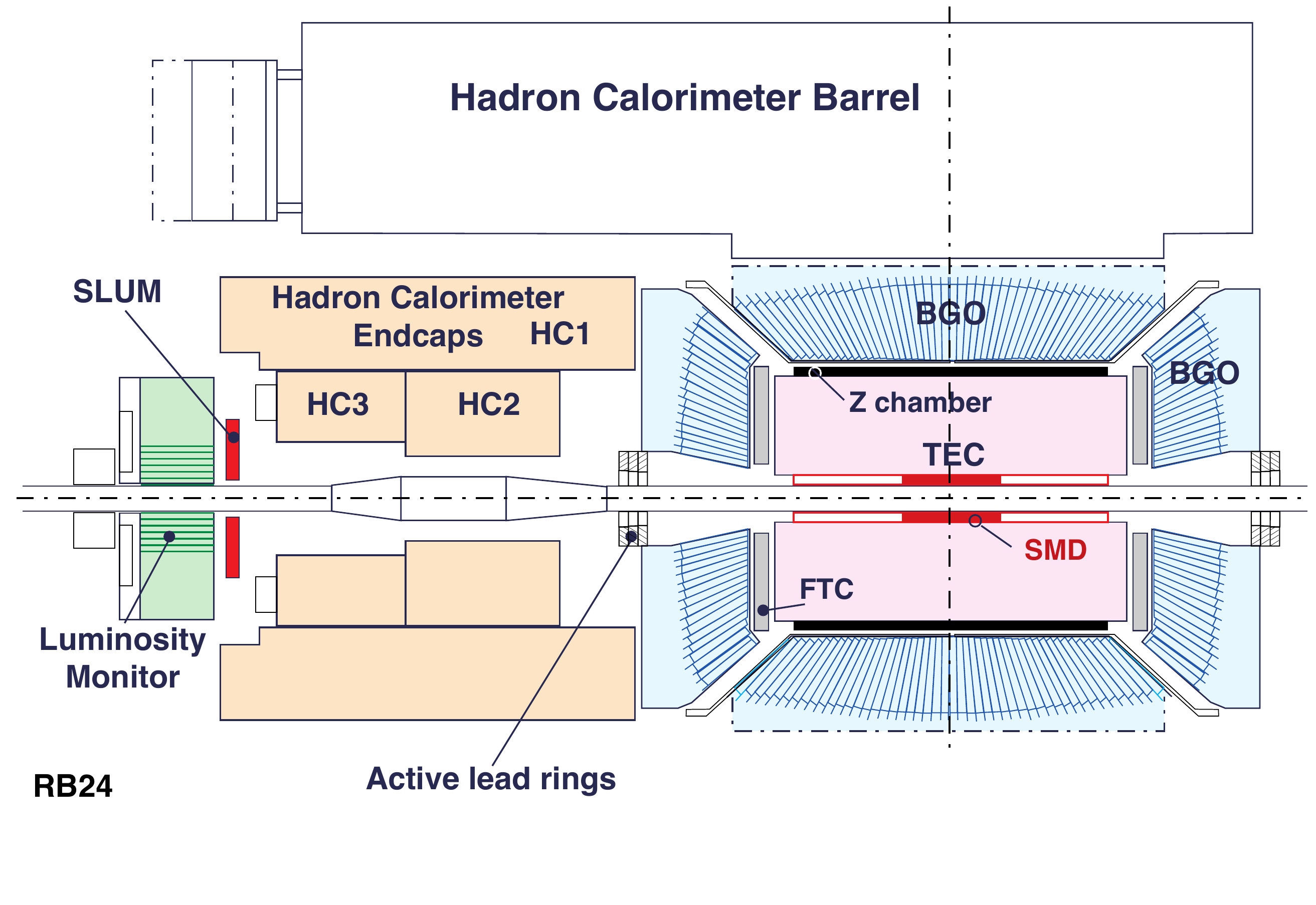}}

\vspace*{-0.5cm}
\caption[luminosity monitors]
{{\sl
Example of a LEP luminosity monitor (the L3 experiment). 
The process to be detected is elastic $e^+e^- \ra e^+e^- $ scattering,
shown on the left, which is seen by a coincidence of an electron in
each of the detectors placed in the forward regions of the detector as
shown on the right. 
In order to determine the solid angle seen by this coincidence in a
way which is independent of the exact location of the collision point
(which is difficult to determine) a set of two different acceptances
is defined for the two arms. The precise definition of the acceptance
is obtained by a precisely machined silicon telescope (SLUM) positioned in
front of the electron calorimeters (luminosity monitor); there is a
symmetric device on the other side of the detector.  
}}
\label{lumimonitor}
\end{figure}

In order to extract a cross-section from the number of events, the luminosity
of the accelerator needs to be determined. ($ N = \cal{L}.\sigma$). This is 
done by measuring at the same time another process with known cross-section,
the elastic scattering $e^+e^- \ra e^+e^-$, known as Bhabha scattering, which
results in two low angle electron and positron.  
To this effect, the LEP experiments were equipped ab initio with low angle 
detectors, to detect these scattered electrons. The precision with which these
detectors can measure this process is determined by the knowledge of the 
solid angle they cover and by the accuracy with which they measure
the angle of the scattered electrons. The initial detectors were able to 
reach a precision of about one percent, but were progressively replaced with 
extremely precisely machined silicon tungsten calorimeters or silicon
trackers (as in figure~\ref{lumimonitor} which allowed 
a determination of the luminosity with an accuracy of $5 10^{-4}$ or better.  
 It took many years of detailed higher order calculations to
achieve a similar precision on the theoretical estimate of the
cross-section 
within this well defined
acceptance. 
 
Measurements of cross-section for a given final state $f\bar{f}$
around the Z pole allows to extract three 
parameters: the position of the peak, the width of the resonance and 
an overall normalisation, that is best obtained from the peak cross-section,
\begin{equation}
\sigma^0_f = \frac{12\pi(\hbar c)^2}{\MMZ} \frac{\Gamma_e \Gamma_f
}{\Gz^2}=\frac{12\pi(\hbar c)^2}{\MMZ} B_e.B_f~. 
\label{sigmafpeak}
\end{equation}

By measuring cross-sections for hadrons, electron pairs, muon pairs and 
tau pairs, one can obtain six numbers:
the mass, the total width, and four other parameters which 
could be four branching ratios, or four partial widths.
A better choice is to use the peak cross-section for hadrons $\s0h$, 
and the ratios of hadrons to the various leptonic partial widths, 
$\Rl \equiv \GH/\GL$.
The Standard Model implies lepton universality, 
and if this is assumed, the number 
of parameters can be reduced to four,
$\MZ,\GZ,\s0h,\Rl$.
The choice of these observables 
to fit the line-shape measurements is dictated by
the fact that they are experimentally uncorrelated, 
both from the point of view
of statistical and systematic errors.

By reporting the expression for $\gamz$
of eq.~\ref{gamz} into the peak cross-section for hadrons, 
eq.~\ref{sigmafpeak}, 
the number of neutrinos can be extracted from quantities 
that are measured at the peak only: 
\begin{equation}
 \nnu = \frac{\gamlep}{\gamnu} 
\cdot \left( \sqrt{{\frac{12\pi \Rl}{\mz^2 \s0h}}} - \Rl -3 \right).
 \label{nnu}
\end{equation}
The sensitivity of $\nnu$ to $\Rl$ is small, as there is a cancellation 
between  the two terms containing this quantity.
As a result the experimental measurement that enters most in the determination
of $\nnu$ is the peak cross-section, as already guessed intuitively from
figure~\ref{Zline}. This observation was made for the first time by G. Feldman~\cite{Feldman:1987}. This explains how quickly the number of neutrinos
was obtained, a few weeks after the start-up of LEP and SLC.

\begin{figure}[htb]

\vspace{-5cm}
\centerline{\includegraphics[width=0.7\textwidth]{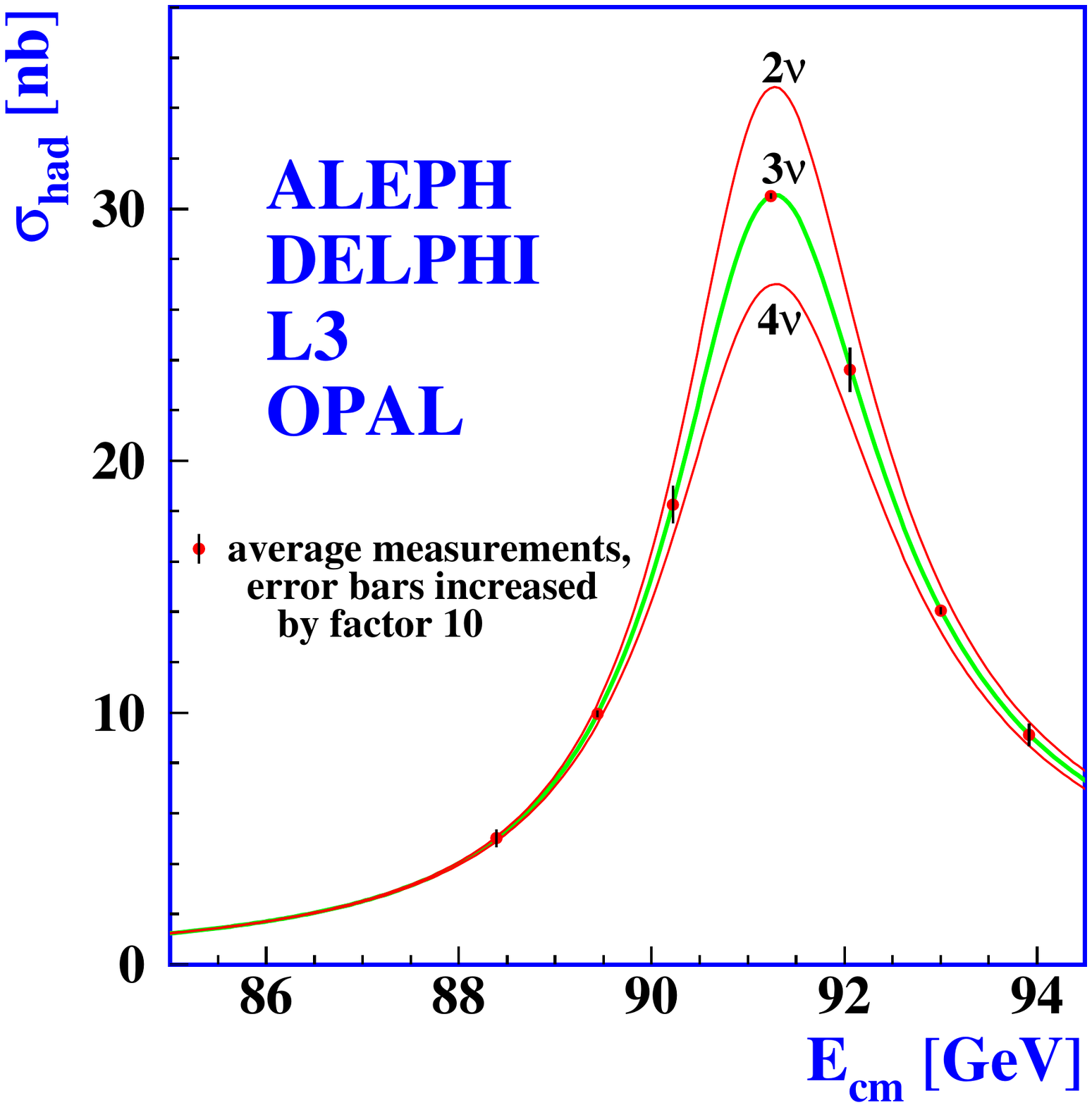}}

\caption[The 1995 hadronic line shape]
{{\sl
The $\epem \ra {\mathrm hadrons}$ cross section
 as a function of center-of-mass energy, as measured 
by  the LEP experiments.  
The curves represent the Standard Model predictions 
for two,  three and four  species of  light neutrinos.
It is clear from this
picture that there is no further light neutrino species with couplings 
identical to the first three.
}}
\label{ALEPHline}
\end{figure}

\newpage

\subsection{The radiative method}

Before the advent of the SLC and LEP, limits on the
number of neutrino generations were placed by experiments at
lower-energy $\epem$ colliders by measuring the cross section of
the process $\epem\rightarrow \nu\nu\gamma$. The ASP, CELLO, MAC, MARKJ,
and VENUS experiments observed a total of 3.9 events above
background~\cite{nunugama-old}, leading to a 95\% CL limit of $\rm{N_\nu < 4.8}$.

This process has a much larger cross section at center-of-mass
energies above the Z mass. The event signal is a photon of an energy consistent with being the recoil against a Z boson. This has been measured at LEP by the ALEPH, DELPHI, L3, and OPAL experiments~\cite{nunugama-Z}. These experiments have observed several thousand such events, and
the combined result is $\rm{N_\nu = 3.00 \pm 0.08}$. The same process has
also been measured by the LEP experiments at much higher
center-of-mass energies, between 130 and 208 GeV, in searches
for new physics~\cite{nunugama-LEPII}. Combined with the lower energy data, the result is $\rm{N_\nu = 2.92 \pm 0.05}$ 

The methods are discussed in~\cite{Trentadue:200667}. The radiative method has sometimes been dubbed 'direct' because the observation of a photon with the proper energy signals the production of an invisible Z boson. By opposition, the Z peak method (which is so far much more precise) was dubbed indirect because the "invisible" Z decays are not counted. This qualification is, again, misleading. In the radiative method, the number of events  $\epem\rightarrow \nu\nu\gamma$ consistent with a Z recoil is proportional to $N_{\nu\nu\gamma}\propto \frac{\Gamma_\nu \Gamma_\nu }{\Gamma_Z}$. Thus a model-independent determination requires an independent measurement of the Z width and its electron partial width. Furthermore the sensitivity to the number of neutrinos is reduced by the fact that $\Gamma_Z$ itself contains the neutrino partial width. This is another example of a 'direct' method which ends up being more model-dependent than the indirect one. 

The radiative method has been proposed in the forthcoming Electroweak factories such as the FCC-ee/TLEP~\cite{Gomez-Ceballos:2013zzn} for a high precision measurement of the neutrino partial width, a quantity sensitive to the mixing of possible heavy right-handed neutrinos with the light active ones.  Many systematic limitations can be eliminated by measuring directly the ratio 
$
R = \frac {e^+ e^- \rightarrow \gamma \nu\nu}{e^+ e^- \rightarrow \gamma \mu\mu}
$
for which, thanks to the high luminosity available at the FCC-ee, a precision of $\Delta{N_\nu} = \pm 0.0008$ i.e. a gain of at least a factor 10, could be achieved.  

\clearpage
\newpage
\vfill\eject

\section{Further developments} 

\subsection{Measurements of the tau-neutrino couplings in tau decays}
\label{tau-couplings-in-tau-decays}

Among the 20 million of visible Z recorded by the LEP collaborations in the seven years 1989-1995, figured in good place one million $\rm{Z\rightarrow \tau^+\tau^-}$. From these large samples a quantum leap in precision was obtained on tau physics, which also benefited from the excellent condition of separation of the tau pair signal from hadronic Z decays at these energies.  Of great interest are the tau life time measurements, the branching ratios into leptons (e, $\mu$) and hadrons, and the $\tau$ mass itself. From these the individual Charged current couplings to $\rm{e-\nu_e,~\mu-\nu_\mu,~\tau-\nu_\tau}$ can be extracted. A good summary is published in~\cite{Schael:2005am}.

In 1991 three of the LEP detector were equipped with silicon vertex detectors and the life time measurements, and in 1992 the BEPC collider provided an improved measurement of the tau mass, and the combination of these results led to greatly improved tests of the universality of Charged Current couplings, as given for instance for the ${\rm \mu-\nu_\mu, vs. ~\tau-\nu_\tau}$ couplings at the Moriond conference in 1993~\cite{Charlton:1993as}:  

\begin{equation}
\frac{g_\tau}{g_\mu} = 0.996 \pm 0.008 
\end{equation}
 
A few years later with the full LEP statistics the precision was down to a few parts permil. Recently these tau measurements have been the object of a yearly update~\cite{lepton-universality-HFLAV-2017}, the latest value being 

\begin{equation}
\frac{g_\tau}{g_\mu} = 1.0010 \pm 0.0015 ~;~  \\   
\frac{g_\tau}{g_e} = 1.0029 \pm 0.0015
\end{equation}	    	
 
\subsection{Measurements of the tau-neutrino couplings in W decays}

\begin{figure}[htb]
\centering
\vspace*{-3cm}
\includegraphics[width=0.8\textwidth]{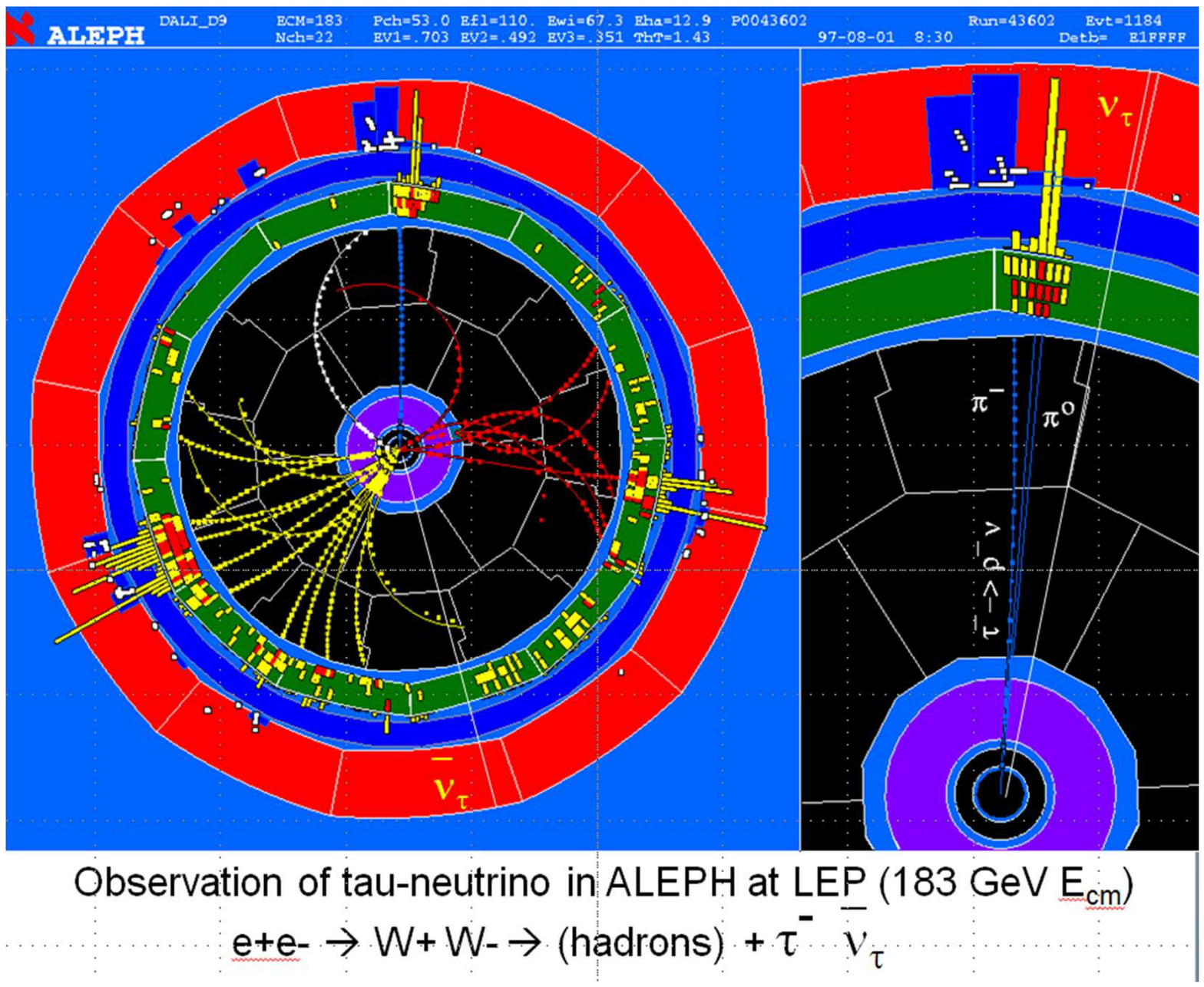}
\vspace*{-5cm}
\caption[A WW to qq tau nu event]
{{\sl
A W pair event recorded in the ALPEH experiment at a centre-of-mass energy of 183GeV (in 1998). One of the W decays in a pair of quark, giving a final state with two jets of hadrons. The other one decays into a $\tau^- \rightarrow \rho^- \bar{\nu_\tau}$ and a tau neutrino which can be reconstructed by constrained kinematics. 
}}
\label{ALEPH-WW-qq-tau-nu_tau}
\end{figure}

At LEP2, tens of thousands of W pair events were detected with a clear event-by-event kinematic reconstruction. An event reconstructed as 
\begin{equation}
\rm{ \epem \rightarrow W^+W^- \rightarrow q\bar{q} \tau^-\bar{\nu_\tau}}  
\end{equation}	    
is shown in Fig.\ref{ALEPH-WW-qq-tau-nu_tau}.

These events allowing determination of all three  W lepton-neutrino couplings at percent level~\cite{Schael:2013ita}.

From the W decays at LEP~\cite{Schael:2013ita}, one can extract the ratios of couplings as follows 
\begin{equation}
\frac{g_\tau}{g_\mu} = 1.070 \pm 0.026 \\
\end{equation}
and from PDG2018, combining the LEP and the Tevatron experiments~\cite{Tanabashi:2018oca}
\begin{equation}
\frac{g_\tau}{g_e} = 1.043 \pm 0.024
\end{equation}	  
which are both less precise than from tau decays and somewhat higher for the tau than for the electron or muon -- a measurement to follow. 


\subsection{Observation of tau neutrino interactions}
\label{DONUT}

\subsubsection{Intent}
In a Fermilab letter of intent from March 1993~\cite{Lundberg:1994wh}, Byron Lundberg and a handful of collaborators (Rameika, Niwa, Reay, Stanton and Paolone) proposed to undertake a 'Measurement of $\tau$ lepton production from the process $\nu_\tau + N \rightarrow \tau$', which would provide 'the first direct evidence for the existence of the $\tau$ neutrino'. 

Interestingly the cover letter "makes it clear that this experiment [...] is a necessary precursor to a sensitive and complete program to measure neutrino mass/mixing here at Fermilab". The experiment was indeed seen as a demonstration that $\tau$ neutrinos could be detected and isolated from backgrounds in the forthcoming NUMI oscillation experiments, which at the time envisaged to place an emulsion neutrino detector in front of MINOS. This motivation was, of course, completely justified, although as it turns out, the experimental demonstration that neutrino oscillations take place mainly among the active flavours would end up being done by MINOS in a completely different way, from charged and neutral current events survival~\cite{Adamson:2011ku}.  

The 'first direct evidence of existence' motivation is contained in the LOI as follows: "Though the top quark has yet to be found, few doubt of its existence. [...] Yet compulsion to produce it and measure it is essential [...]. The existence of a tau neutrino partner for the tau was assumed and indirectly confirmed by the tau decay spectrum. (...) Like with the top quark, few doubt that the tau lepton does have its own neutrino, however direct observation of the   $\nu_\tau \rightarrow \tau$ charged current interaction would confirm its existence".  

\subsubsection{Discussion}
It is evident that the writers of these lines were not aware of the details reported in the earlier sections of this article, by which the existence of the tau neutrino had been duly and rigorously established in the 1980's -- 1986 at the latest. Ignored is $\tau \rightarrow \pi \nu$ decay, the measurement of the tau lifetime, of the observation of $W\rightarrow \tau \nu_\tau$ or the findings of Fermilab experiment E531, and indeed there is in the LOI no reference at all to the earlier existing information concerning the tau neutrino. 

The analogy with the top quark is particularly telling (of course it could simply be done to catch the ears of a committee more occupied in 1993 with the top quark search than with tau neutrinos): the postulated existence of the top quark was purely theoretical, being a consequence of the isospin symmetry of the Standard Model, the b  quark (resp $\tau$ lepton) having been found with isospin -1/2, a particle with isospin +1/2 had to exist. Nevertheless many "topless models" existed in the late 80's to early 90's~\cite{topless}\cite{Barger:128662}\cite{Grigorian:138135}\cite{Truini:147514}\cite{Grigorian:183754}\cite{Pakvasa:206725}\cite{Fujiwara:218609}, showing that the existence of the top quark was not only 'indirect' but more importantly and meaningfully, 'uncertain'. Even in 1994, after the top quark effect had been seen in the self-energy radiative corrections e.g. to the Z width, allowing an 'indirect' prediction of its mass, this was still a model-dependent prediction, specifically assuming that the minimal Standard Model with three families, a single Higgs boson and {\sl nothing else} is at play. 

Of course there is no such parallel between the top and the tau neutrino: nobody had ever seen the production of the top quark in 1993, while hundreds of thousands of tau neutrinos had been seen to be produced in tau decays, as well as in on-shell and off-shell W decays, its existence as distinct from $\nu_e$ and $\nu_\mu$ established, and its properties duly measured.  While the existence of the top quark was not only 'indirect' but arguably 'uncertain', the status of the tau neutrino was 'established' and even 'measured' -- two very, very different levels of status, confused by the thin and ill-defined link of using the word 'indirect'.    

Why was there such  a gap in appreciation between the neutrino beam community and the collider community? As I personally know both of them, I can say that at collider a neutrino is a missing 4-momentum (less $p_z$ for a hadron collider) of invariant mass consistent with 0, and which has, by definition, the same lepton flavour (if one cares) as the lepton it appears with. When a neutrino beam physicist talks of a 'muon neutrino', he has in mind a neutrino interaction producing a muon, ($CC_{\nu_\mu}$), same with an 'electron neutrino' producing an electron, and of course a 'tau neutrino' will be a neutrino interaction producing a tau. The two definitions are of course completely equivalent scientifically, but not visually; both encompass a large amount of underlying demonstration of the fact that the neutrino states are essentially orthogonal (oscillations happen at much longer time scale). Thus the neutrino part of the particle physics world had been longing for nearly twenty years to {\bf see} a 'tau neutrino' -- that is, its interaction. Indeed, to see it would be a feast... especially since it is very difficult to achieve! 


\begin{figure}[htb]
\centering
\vspace*{-1cm}
\includegraphics[width=0.8\textwidth]{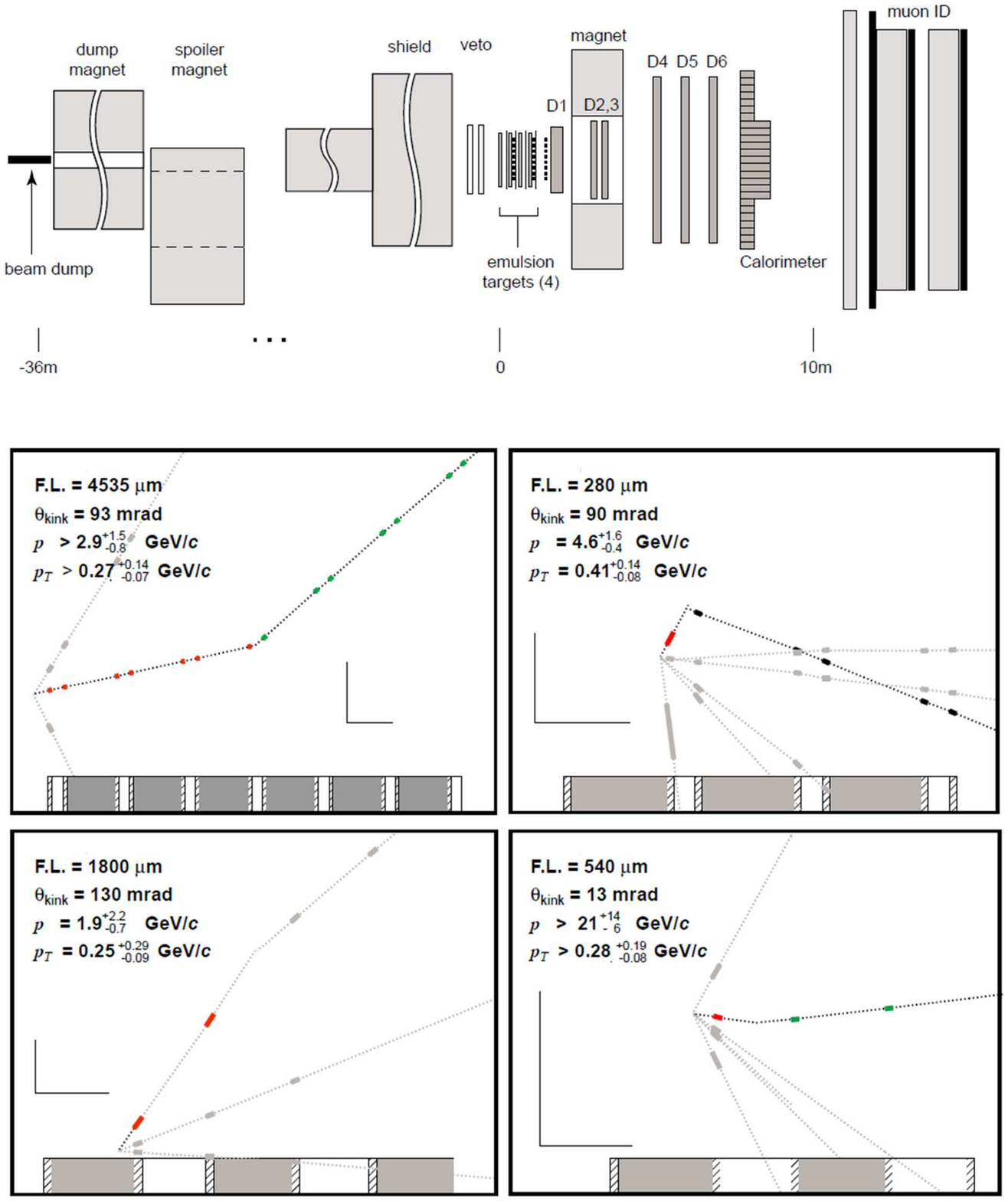}
\vspace*{-2cm}
\caption[DONUT]
{{\sl The DONUT experiment. Top: overall set-up, with the beam coming from the left. In succession, the beam dump; spoiler magnet to remove sweep muons aside; shield and veto counter; the emulsion targets followed by the magnetic spectrometer, a calorimeter and muon filter.  Bottom: four tau neutrino events of the first publication.
}}
\label{DONUT-figure}
\end{figure}

\subsubsection{The DONUT experiment}
The DONUT experiment principle~\cite{Kodama:2000mp} is to produce a beam enriched in tau neutrinos by using a beam dump. The highest energy (800 GeV) protons ever available for fixed target experiments are delivered from the Tevatron on a tungsten beam dump, where $\rm D_s$ mesons are produced and decay into $\rm{D_s\rightarrow \tau + \bar{\nu}_\tau }$, followed by the tau decay into a tau neutrino. The tau neutrino from the tau decay is more energetic, thus better focused and of higher cross-section than the tau antineutrino, so its interactions are expected to dominate. The beam dump ensures that most pions and kaons are absorbed. Thanks to a magnetized iron muon sweeping magnet and to a shield, the detector is situated remarkably close (38 m) from the beam dump. Altogether the neutrino interactions are expected to consist of 95\% electron and muon (anti)neutrinos and 5\% tau neutrinos. This is a striking difference with the E531 experiment mentioned above, which was situated in a regular neutrino beam in which the tau neutrino component is proportionally more than 3000 times smaller. 

The overall set-up is illustrated in Fig.\ref{DONUT-figure}. It is notable that the experiment made first use of the Emulsion Cloud Chamber (ECC) where metal plates were interleaved with two planes of emulsions separated by 0.8mm of plastic, allowing both a higher target mass and precision determination of the direction of tracks. 

The first publication~\cite{Kodama:2000mp} 'observation of tau neutrino interactions' reported four such events, also pictured in Fig.\ref{DONUT-figure}, out of 203 neutrino interactions located in the emulsion. 

The paper was careful to stress that 
"It should be noted that since the neutrino flux had only an estimated 5\%  component, the possibility that the ${\nu_\tau}$ is a superposition of ${\nu_\mu}$ and  ${\nu_e}$ cannot be eliminated using the results of this experiment. Results from other experiments (quoting the E531 results and more recent ones from Chorus and NOMAD) which were sensitive to $\tau$ leptons, show that the direct coupling of ${\nu_\mu}$ to $\tau$ is very small (2$\times 10^{-4}$). The upper limit (90\% CL) for  ${\nu_e}$ to  $\tau$  is much larger, $1.1\times 10^{-2}$ (90\% CL). Assuming this upper limit, the estimated number of $\tau$ events from this hypothetical source is $0.27\pm 0.09$ (90\% CL)". 

The probability that four events came from the background was 4$\times 10^{-4}$ -- equivalent to a 3.5 sigma significance, perfectly sufficient to convince everyone that DONUT had indeed seen the first tau neutrino interactions with matter ever recorded and identified; and the events were beautiful. "Direct observation of $\nu_\tau$, that is, interactions resulting from neutrinos produced in tau lepton decay", had been made. "That is", by definition.    

With great enthusiasm the Fermilab News announced that "DONUT Finds Missing Puzzle Piece"~\cite{DONUTferminews}. This was of course over-enthusiastic, this piece of the Standard Model puzzle having been found many years before. Just to be clear: DONUT did not discover the tau neutrino, and the collaboration never made this claim in a scientific paper. However DONUT had satisfied a long-standing longing of the neutrino community, and fullfilled its mission by demonstrating the feasibility of observing tau neutrino interactions, in view of possible neutrino oscillation experiments.

\section{Acknowledgements}
It is a pleasure to thank the organizers of the conference for their kind invitation to present this largely forgotten piece of neutrino history. This research, based on work originally performed in the late 80's, has benefited from extensive discussions with F. Dydak, M. Davier, T. Pich and Jack Steinberger, and more recently G. Feldman, S. Petcov and V. Paolone, whom I gratefully acknowledge. I will be happy to complete this document with further historical comments, or with facts that I may have easily overlooked.

\newpage

\newpage
\vfill\eject
\medskip
 
\bibliographystyle{unsrt}
\bibliography{bib/nutau}

\begin{thebibliography}{10}

\bibitem{Perl:1975bf}
Martin~L. Perl et~al.
\newblock {Evidence for Anomalous Lepton Production in e+ - e- Annihilation}.
\newblock {\em Phys. Rev. Lett.}, 35:1489--1492, 1975.

\bibitem{Feldman:1976fm}
G.~J. Feldman et~al.
\newblock {Inclusive Anomalous Muon Production in e+ e- Annihilation}.
\newblock {\em Phys. Rev. Lett.}, 38:117, 1977.
\newblock [Erratum: Phys. Rev. Lett.38,576(1977)].

\bibitem{Perl:1977se}
Martin~L. Perl et~al.
\newblock {Properties of the Proposed tau Charged Lepton}.
\newblock {\em Phys. Lett.}, 70B:487--490, 1977.

\bibitem{Bacino:1979fz}
W.~Bacino et~al.
\newblock {On the Nature of the tau Tau-neutrino W Coupling}.
\newblock {\em Phys. Rev. Lett.}, 42:749, 1979.

\bibitem{Kirkby:1979pv}
Jasper Kirkby.
\newblock {Review of $e^+ e^-$ Reactions in the Energy Range 3-{GeV} to
  9-{GeV}}.
\newblock In {\em {Proceedings: International Symposium on Lepton and Photon
  Interactions at High Energies, Batavia, Ill., Aug 23-29, 1979}}, 1979.
\newblock
  {http://www-public.slac.stanford.edu/sciDoc/docMeta.aspx?slacPubNumber=SLAC-PUB-2419}.

\bibitem{BarbaroGaltieri:1977ti}
Angela Barbaro-Galtieri et~al.
\newblock {Electron-Muon and electron-Hadron Production in e+ e- Collisions}.
\newblock {\em Phys. Rev. Lett.}, 39:1058, 1977.

\bibitem{Brandelik:1977xz}
R.~Brandelik et~al.
\newblock {Measurements of tau Decay Modes and a Precise Determination of the
  Mass}.
\newblock {\em Phys. Lett.}, 73B:109--114, 1978.

\bibitem{Blocker:1981mc}
C.~A. Blocker et~al.
\newblock {A Study of the Decay $\tau^- \to \pi^- \nu_\tau$}.
\newblock {\em Phys. Lett.}, 109B:119--123, 1982.

\bibitem{Feldman:1981nv}
Gary~J. Feldman.
\newblock {The lepton spectrum, SLAC-PUB-2839}.
\newblock {\em AIP Conf. Proc.}, 81:280--302, 1982.

\bibitem{Cnops:1977zc}
A.~M. Cnops et~al.
\newblock {Experimental Limits on Heavy Lepton Production by Neutrinos}.
\newblock {\em Phys. Rev. Lett.}, 40:144--146, 1978.

\bibitem{Fritze:1980un}
P.~Fritze et~al.
\newblock {Further Study of the Prompt Neutrino Flux From 400-{GeV} Proton -
  Nucleus Collisions Using {BEBC}}.
\newblock {\em Phys. Lett.}, 96B:427--434, 1980.

\bibitem{Feldman:1981md}
G.~J. Feldman et~al.
\newblock {Measurement of the $\tau$ Lifetime}.
\newblock {\em Phys. Rev. Lett.}, 48:66, 1982.

\bibitem{Roos:1982sd}
M.~Roos et~al.
\newblock {Review of Particle Properties. Particle Data Group}.
\newblock {\em Phys. Lett.}, 111B:1--294, 1982.

\bibitem{PDGarchives}
{Particle Data Group, Previous Editions (\& Errata) 1957-2017}.
\newblock \url{http://pdg.lbl.gov/rpp-archive/}, 2018.

\bibitem{Vuillemin:1985tc}
V.~Vuillemin.
\newblock {W+- and Z0 Production in the UA1 experiment at the CERN proton -
  antiproton Collider}.
\newblock {\em Annals N. Y. Acad. Sci.}, 461:99, 1986.

\bibitem{Albajar:1986fn}
C.~Albajar et~al.
\newblock {Events with Large Missing Transverse Energy at the CERN Collider: W
  ---> tau-neutrino Decay and Test of tau - mu - e Universality at Q**2 =
  m(w**2)}.
\newblock {\em Phys. Lett.}, B185:233--240, 1987.
\newblock [Addendum: Phys. Lett.B191,462(1987)].

\bibitem{Albajar:1988ka}
C.~Albajar et~al.
\newblock {Studies of Intermediate Vector Boson Production and Decay in UA1 at
  the CERN Proton - Antiproton Collider}.
\newblock {\em Z. Phys.}, C44:15--61, 1989.

\bibitem{Ushida:1986zn}
N.~Ushida et~al.
\newblock {Limits to Muon-neutrino, electron-neutrino ---> tau-neutrino
  Oscillations and Muon-neutrino, electron-neutrino ---> tau- Direct Coupling}.
\newblock {\em Phys. Rev. Lett.}, 57:2897--2900, 1986.

\bibitem{Denegri:1989if}
D.~Denegri, B.~Sadoulet, and M.~Spiro.
\newblock {The Number of Neutrino Species, CERN-EP-89-72}.
\newblock {\em Rev. Mod. Phys.}, 62:1--42, 1990.

\bibitem{Abrams:1989yk}
G.~S. Abrams et~al.
\newblock {Measurements of $Z$ Boson Resonance Parameters in $e^{+} e^{-}$
  Annihilation}.
\newblock {\em Phys. Rev. Lett.}, 63:2173, 1989.

\bibitem{Adeva:1989mn}
B.~Adeva et~al.
\newblock {A Determination of the Properties of the Neutral Intermediate Vector
  Boson $Z^0$}.
\newblock {\em Phys. Lett.}, B231:509, 1989.

\bibitem{Decamp:1989tu}
D.~Decamp et~al.
\newblock {Determination of the Number of Light Neutrino Species}.
\newblock {\em Phys. Lett.}, B231:519--529, 1989.

\bibitem{Akrawy:1989pi}
M.~Z. Akrawy et~al.
\newblock {Measurement of the $Z^0$ Mass and Width with the OPAL Detector at
  LEP}.
\newblock {\em Phys. Lett.}, B231:530--538, 1989.

\bibitem{Aarnio:1989tv}
P.~Aarnio et~al.
\newblock {Measurement of the Mass and Width of the $Z^0$ Particle from Multi -
  Hadronic Final States Produced in $e^{+} e^{-}$ Annihilations}.
\newblock {\em Phys. Lett.}, B231:539--547, 1989.

\bibitem{blondel1991}
A.~Blondel.
\newblock {La première expérience d'ALEPH au LEP}.
\newblock In {\em Mémoires de l'Académie des Sciences, Belles-Lettres et Arts
  de Lyon, Lyon 1992, 4 rue Adolph-Max, 69005 Lyon}, volume troisième série,
  tome quarante sixième, pages 202--205, 1992.

\bibitem{ALEPH:2005ab}
S.~Schael et~al.
\newblock {Precision electroweak measurements on the $Z$ resonance}.
\newblock {\em Phys.Rept.}, 427:257--454, 2006.

\bibitem{Charlton:1993as}
David~G. Charlton.
\newblock {Tau lepton lifetime and decay branching ratio measurements at LEP}.
\newblock In {\em {1993 electroweak interactions and unified theories.
  Proceedings, Leptonic Session of the 28th Rencontres de Moriond, Les Arcs,
  France, March 13-20, 1993}}, pages 349--356, 1993.
\newblock {https://cds.cern.ch/record/284476}.

\bibitem{lepton-universality-HFLAV-2017}
{HFLAV tau spring 2017 report, Tests of lepton universality}.
\newblock
  \url{http://www.slac.stanford.edu/xorg/hflav/tau/spring-2017/lepton-univ.html},
  2017.

\bibitem{Schael:2013ita}
S.~Schael et~al.
\newblock {Electroweak Measurements in Electron-Positron Collisions at
  W-Boson-Pair Energies at LEP}.
\newblock {\em Phys. Rept.}, 532:119--244, 2013.

\bibitem{Nakamura:1999dp}
M.~Nakamura.
\newblock {Result from DONUT: Direct observation of nu/tau interaction}.
\newblock {\em Nucl. Phys. Proc. Suppl.}, 77:259--264, 1999.
\newblock [,259(1999)].

\bibitem{Kodama:2000mp}
K.~Kodama et~al.
\newblock {Observation of tau neutrino interactions}.
\newblock {\em Phys. Lett.}, B504:218--224, 2001.

\bibitem{Kodama:2007aa}
K.~Kodama et~al.
\newblock {Final tau-neutrino results from the DONuT experiment}.
\newblock {\em Phys. Rev.}, D78:052002, 2008.

\bibitem{Feldman:1992vk}
Gary~J. Feldman.
\newblock {The Discovery of the tau, 1975-1977: A Tale of three papers; {\sl
  in} The Third family and the physics of flavor: Proceedings, 20th SLAC summer
  institute on particle physics (SSI 92), 13-24 Jul 1992}.
\newblock {\em Conf. Proc.}, C9207131:631--646, 1992.

\bibitem{Tsai:1971vv}
Yung-Su Tsai.
\newblock {Decay Correlations of Heavy Leptons in e+ e- ---> Lepton+ Lepton-}.
\newblock {\em Phys. Rev.}, D4:2821, 1971.
\newblock [Erratum: Phys. Rev.D13,771(1976)].

\bibitem{Perl:1976rz}
Martin~L. Perl et~al.
\newblock {Properties of Anomalous e mu Events Produced in e+ e- Annihilation}.
\newblock {\em Phys. Lett.}, 63B:466, 1976.

\bibitem{Danby:1962nd}
G.~Danby, J.~M. Gaillard, Konstantin~A. Goulianos, L.~M. Lederman, Nari~B.
  Mistry, M.~Schwartz, and J.~Steinberger.
\newblock {Observation of High-Energy Neutrino Reactions and the Existence of
  Two Kinds of Neutrinos}.
\newblock {\em Phys. Rev. Lett.}, 9:36--44, 1962.

\bibitem{Fernandez:1983az}
E.~Fernandez et~al.
\newblock {Lifetime of Particles Containing B Quarks}.
\newblock {\em Phys. Rev. Lett.}, 51:1022, 1983.

\bibitem{Lockyer:1983ev}
N.~Lockyer et~al.
\newblock {Measurement of the Lifetime of Bottom Hadrons}.
\newblock {\em Phys. Rev. Lett.}, 51:1316, 1983.

\bibitem{petcov2018}
{S. Petcov reminded me of that discussion, but neither of us could find a
  reference to it}.

\bibitem{Brandelik:1980ga}
R.~Brandelik et~al.
\newblock {Production and Properties of the $\tau$ Lepton in $e^+ e^-$
  Annihilation at c.m. Energies From 12-{GeV} to 31.6-{GeV}}.
\newblock {\em Phys. Lett.}, 92B:199--205, 1980.

\bibitem{Jaros:1983uq}
J.~Jaros et~al.
\newblock {Precise Measurement of the Tau Lifetime}.
\newblock {\em Phys. Rev. Lett.}, 51:955, 1983.

\bibitem{AguilarBenitez:1986fu}
M.~Aguilar-Benitez et~al.
\newblock {Review of Particle Properties. Particle Data Group}.
\newblock {\em Phys. Lett.}, 170B:1--350, 1986.

\bibitem{Ellis:1984sz}
John~R. Ellis and Marc Sher.
\newblock {Is Supersymmetry found?}
\newblock {\em Phys. Lett.}, 148B:309--316, 1984.

\bibitem{Ellis:1985xw}
S.~D. Ellis, R.~Kleiss, and W.~James Stirling.
\newblock {The Standard Model and missing E(T) or the many Roads to Paradise,
  CERN-TH-4170/85}.
\newblock In {\em {5th Topical Workshop on Proton Antiproton Collider Physics
  St.Vincent, Italy, February 25-March 2, 1985}}, 1985.

\bibitem{Dilella2016}
{L. Di Lella, The Altarelli Cocktail and other memories, {\sl Guido Altarelli
  Memorial Symposium} }.
\newblock \url{https://indico.cern.ch/event/493632/}, 2016.

\bibitem{Barish:1987nj}
B.~C. Barish and R.~Stroynowski.
\newblock {The Physics of the tau Lepton}.
\newblock {\em Phys. Rept.}, 157:1, 1988.

\bibitem{Dydak:171238}
Friedrich Dydak.
\newblock {Experimental Tests of the Electroweak Theory}.
\newblock (CERN-EP-86-121):67 p, Sep 1986.
\newblock {http://cds.cern.ch/record/171238}.

\bibitem{Pich:1989qk}
A.~Pich.
\newblock {Tau Physics: Present status and Future Prospects}.
\newblock In {\em {Physics at LEP. Proceedings, 17th International Meeting on
  Fundamental Physics, Lekeitio, Spain, April 23-29, 1989}}, pages 0323--362,
  1989.

\bibitem{Winter:1989mj}
Klaus Winter.
\newblock {Detection of the tau-neutrino, CERN-EP/89-182}.
\newblock In {\em {Proceedings, Tau Lepton Physics (TAU 90): Orsay, France,
  September 24-27, 1990}}.

\bibitem{Weinberg:1972kfs}
Steven Weinberg.
\newblock {\em {Gravitation and Cosmology}}.
\newblock John Wiley and Sons, New York, 1972.
\newblock {http://www-spires.fnal.gov/spires/find/books/www?cl=QC6.W431}.

\bibitem{Ellis:875312}
Jonathan~Richard Ellis.
\newblock {e+e- Physics at LEP Energies: Zedology}.
\newblock (LEP-Summer-Study-1-14-1), 1979.
\newblock {https://cds.cern.ch/record/875312}.

\bibitem{Abrams:1989aw}
G.~S. Abrams et~al.
\newblock {Initial Measurements of Z Boson Resonance Parameters in e+ e-
  Annihilation}.
\newblock {\em Phys. Rev. Lett.}, 63:724, 1989.

\bibitem{Feldman:1987}
Gary~J. Feldman.
\newblock {On the Possibility of Measuring the Number of Neutrino Species to a
  Precision of 1/2 Species with Only 2000 Z Events, {\sl in proc. 3d MARK II
  workshop on SLC physics}, SLAC-R-315}.
\newblock Technical report, 1987.
\newblock {http://slac.stanford.edu/pubs/slacreports/reports04/slac-r-315.pdf}.

\bibitem{nunugama-old}
{ VENUS: K. Abe et al., Phys. Lett. B232, 431 (1989); ASP: C. Hearty et al.,
  Phys. Rev. D39, 3207 (1989); CELLO: H.J. Behrend et al., Phys. Lett. B215,
  186 (1988); MAC: W.T. Ford et al., Phys. Rev. D33, 3472 (1986); MARK J: H.
  Wu, Ph.D. Thesis, Univ. Hamburg (1986)}.

\bibitem{nunugama-Z}
{ L3: M. Acciarri et al., Phys. Lett. B431, 199 (1998); DELPHI: P. Abreu et
  al., Z. Phys. C74, 577 (1997); OPAL: R. Akers et al., Z. Phys. C65, 47
  (1995); ALEPH: D. Buskulic et al., Phys. Lett. B313, 520 (1993) }.

\bibitem{nunugama-LEPII}
{ DELPHI: J. Abdallah et al., Eur. Phys. J. C38, 395 (2005); L3: P. Achard et
  al., Phys. Lett. B587, 16 (2004); ALEPH: A. Heister et al., Eur. Phys. J.
  C28, 1 (2003); OPAL: G. Abbiendi et al., Eur. Phys. J. C18, 253 (2000). }.

\bibitem{Trentadue:200667}
L~G Trentadue, Guido Barbiellini, X~Berdugo, G~Bonvicini, P~Colas, L~Mirabito,
  C~Dionisi, D~A Karlen, Frank~L Linde, C~Luci, C~Maña, C~Matteuzzi,
  O~Nicrosini, R~Ragazzon, A~D Schaile, and F~Scuri.
\newblock {Neutrino counting}.
\newblock (CERN-TH-5528-89):42 p, Sep 1989.
\newblock {http://cds.cern.ch/record/200667}.

\bibitem{Gomez-Ceballos:2013zzn}
M.~Bicer et~al.
\newblock {First Look at the Physics Case of TLEP, arxiv:1308.6176}.
\newblock {\em JHEP}, 01:164, 2014.

\bibitem{Schael:2005am}
S.~Schael et~al.
\newblock {Branching ratios and spectral functions of tau decays: Final ALEPH
  measurements and physics implications}.
\newblock {\em Phys. Rept.}, 421:191--284, 2005.

\bibitem{Tanabashi:2018oca}
M.~Tanabashi et~al.
\newblock {Review of Particle Physics}.
\newblock {\em Phys. Rev.}, D98(3):030001, 2018.

\bibitem{Lundberg:1994wh}
B.~Lundberg et~al.
\newblock {Measurement of tau lepton production from the process tau-neutrino +
  N -> tau, Fermilab-Proposal-0872}.
\newblock 1994.

\bibitem{Adamson:2011ku}
P.~Adamson et~al.
\newblock {Active to sterile neutrino mixing limits from neutral-current
  interactions in MINOS}.
\newblock {\em Phys. Rev. Lett.}, 107:011802, 2011.

\bibitem{topless}
{A search in the cern library yields six entries with 'topless model' in the
  title, I apologize for giving only those below, recognizing that I have no
  idea how many more papers studied the question of having no top quark while
  the b quark was measured to have weak isospin of -1/2.}

\bibitem{Barger:128662}
V~Barger, W~Y Keung, and R~J~N Phillips.
\newblock {Study of b-couplings in topless and standard quark models}.
\newblock Technical Report DOE-ER-00881-190, Wisconsin Univ., Madison, WI, Mar
  1981.
\newblock {https://cds.cern.ch/record/128662}.

\bibitem{Grigorian:138135}
S~S Grigorian and S~F Sultanov.
\newblock {B-meson decays in left-right symmetric topless models}.
\newblock Technical Report IFVE-82-61. IFVE-OTF-82-61, Akad. Nauk Kazakhskoj.
  Inst. Fiz. Vysok. Energy, Alma-Ata, Mar 1982.
\newblock {https://cds.cern.ch/record/138135}.

\bibitem{Truini:147514}
P~Truini, Lawrence~Christian Biedenharn, and B~F~L Ward.
\newblock {Kaon yield per B decay: a test for topless models?}
\newblock Technical Report SLAC-PUB-3103, SLAC, Stanford, CA, May 1983.
\newblock {https://cds.cern.ch/record/147514}.

\bibitem{Grigorian:183754}
S~S Grigorian and S~F Sultanov.
\newblock {Charge asymmetry in $e^+e^- \rightarrow b\overline{b}$ against
  topless models}.
\newblock Technical Report IFVE-87-140. IFVE-OTF-87-140, Akad. Nauk Kazakhskoj.
  Inst. Fiz. Vysok. Energy, Alma-Ata, Aug 1987.
\newblock {https://cds.cern.ch/record/183754}.

\bibitem{Pakvasa:206725}
S~Pakvasa, D~P Roy, and S~Uma-Sankar.
\newblock {Phenomenological analysis of a topless left-right model}.
\newblock Technical Report TIFR-TH-90-11. UH-511-694, Tata Inst. Fundam. Res.,
  Bombay, Mar 1990.
\newblock {https://cds.cern.ch/record/206725}.

\bibitem{Fujiwara:218609}
T~Fujiwara and S~Kitakado.
\newblock {Topless electroweak model as anomalous gauge theory}.
\newblock Technical Report IU-91-1, Ibaraki Univ., Ibaraki, Feb 1991.
\newblock {https://cds.cern.ch/record/218609}.

\bibitem{DONUTferminews}
{DONUT Finds Missing Puzzle Piece}.
\newblock \url{https://www.fnal.gov/pub/ferminews/ferminews00-08-04/p1.html},
  2000.

\end{thebibliography}







\end{document}